# Near-infrared lensless holographic microscopy on a visible sensor enables label-free high-throughput imaging in strong scattering

Emilia Wdowiak,[a*] Piotr Arcab,[a] Mikołaj Rogalski,[a] Anna Chwastowicz,[b,c] Paweł Matryba,[b,c] Małgorzata Lenarcik,[d,e] Julianna Winnik,[a] Piotr Zdańkowski,[a] Maciej Trusiak[a**]

[a] Warsaw University of Technology, Institute of Micromechanics and Photonics, 8 Sw. A. Boboli St., 02-525 Warsaw, Poland

[b] Laboratory of Neurobiology, Nencki Institute of Experimental Biology of Polish Academy of Sciences, Warsaw, Poland

[c] Department of Immunology, Medical University of Warsaw, Warsaw, Poland

[d] Department of Pathology, Maria Sklodowska-Curie National Research Institute of Oncology, Warsaw, Poland

[e] Department of Gastroenterology, Hepatology and Clinical Oncology, Centre of Postgraduate Medical Education, Warsaw, Poland

**Abstract.** Lensless digital holographic microscopy (LDHM) relies on interference between an unscattered reference wave and a weakly scattered object wave - an assumption that rapidly fails in turbid samples under multiple scattering. To overcome this limitation, we present near-infrared LDHM (NIR-LDHM), a in-line holographic platform that operates up to the silicon cutoff (~1100 nm) using a conventional board-level CMOS sensor designed for visible (VIS) imaging. Using tissue-mimicking milk scattering layers and calibrated resolution targets, we quantify reconstruction performance versus wavelength, scattering strength, and sample-sensor distance. NIR-LDHM maintains resolvable features through scattering layers up to ~1.4 mm, whereas visible regime fails to resolve features below ~350 μm. Importantly, despite a detector quantum efficiency of only 0.19% at 1100 nm, robust reconstructions are obtained under low-photon-budget conditions. We further identify advantageous mechanism by which increasing the sample-sensor distance from ~3 to 12 mm improves lateral resolution by twofold under strong scattering. Finally, we demonstrate wide-field, label-free amplitude-phase imaging of uncleared mouse tissues, resolving internal structure in brain and liver slices up to ~250 μm and ~60 μm, respectively. By extending lensless complex-field microscopy into strongly scattering regimes with minimal hardware changes, this work has relevance to computational imaging through complex media and biophotonics.

*Emilia Wdowiak**,** E-mail: emilia.wdowiak.dokt@pw.edu.pl
**Maciej Trusiak**,** E-mail: maciej.trusiak@pw.edu.pl

# 1 Introduction

Optical microscopy plays a crucial role in scientific discoveries in biology, biomedicine and biophotonics. Its classical realization provides amplitude contrast driven by sample induced spatial variations in the intensity of light. Quantitative phase imaging (QPI) has become a key modality for label-free characterization of cells and tissues, providing nanometer-scale sensitivity to optical thickness fluctuations (physical thickness and refractive index - RI) without exogenous contrast agents[1–4]. By measuring refraction-related phase delay, complementary to the classically available absorption-related intensity contrast, QPI yields additional label-free, intrinsic biophysical readouts, such as dry-mass density, cell growth and dynamics, and microstructural tissue composition that are directly relevant to both basic biology and clinical diagnostics[1–4]. Most mature modern microscopy implementations are built around high-numerical aperture (NA) microscope objectives embedded in interferometric architectures[1–4]. This inherently limits field of view (FOV) and depth of field (DOF), enforcing a trade-off between resolution and sample imaging area. It also leads to bulky, alignment-sensitive, and expensive instruments that are difficult to scale to mesoscale tissue imaging or deploy outside specialized optics laboratories. These constraints have motivated intense interest in lensless digital holographic microscopy (LDHM)[5,6], based on seminal work by Dennis Gabor[5]. In LDHM with maximized FOV sample is placed directly in front of a board-level sensor and the complex field - both amplitude and phase contrast - is reconstructed numerically upon propagation from a single or multiple holograms[7,8]. By eliminating imaging optics, LDHM offers centimeter-scale FOVs, extended DOF and robust common-path in-line geometry in a compact, low-cost platform[6–8].

The classical Gabor in-line holographic model is fundamentally rooted in a weak-scattering regime[5]. Effectively, single-scattering is considered: the object is treated as a small perturbation on a dominant unscattered reference wave, and multiple scattering is neglected[5]. This approximation generally holds for isolated cells and mainly transparent structures like very thin tissue slices, but it breaks down in optically thick, heterogeneous and multiple scattering specimens. Yet this is precisely the regime of tissue slices imaging, central to biomedicine and diagnostics, especially challenging within the label-free imaging methods. Vibratome-cut slices of brain, tumors and other organs routinely extend from tens to hundreds of micrometers depths and exhibit substantial multiple scattering due to strong RI fluctuations at cellular and subcellular levels[9,10]. Digital holographic microscopy and tomography have been successfully applied only to standard histological thicknesses (≈ 3-15 μm)[1–4]. Thicker slices, although more practical and information-rich, require special treatments under contemporary development, predominantly with lens-based FOV-limited techniques, including (1) numerical multiple scattering approaches[11–16] and/or (2) physically lowered coherence of illumination and common path configurations[9,17–20] to enable deeper penetration up to ~200 μm.

Near-infrared (NIR) holographic tomography has been implemented to scattering phantom imaging[21] and 15 μm thick colon slices examination[22], showing great potential yet still underexplored for high-throughput deep tissue imaging, partly due to the need for NIR-optimized optical components, which are challenging to align and expensive. Importantly, in soft tissues and tissue-like phantoms, scattering anisotropy and attenuation generally decrease with increasing wavelength, while endogenous absorption is reduced in NIR compared with the visible (VIS) range, resulting in a higher fraction of signal photons reaching the detector. This trend follows from the fact that scattering is strongest on inhomogeneities comparable in size to the illumination wavelength; therefore, increasing the wavelength reduces scattering effects. A similar advantage is exploited in multiphoton infrared imaging[23]; however, to date, NIR has not been translated to LDHM. In widefield LDHM configuration, for thicker sections or imaging through tissue-like diffusers, the Born approximation fails, twin-image artifacts are amplified, and the recovered phase no longer faithfully reflects morphology. Initial works on this topic show that optical clearing allows for LDHM large space-bandwidth product imaging of mouse brain slices up to 200 μm[24] and 500 μm[25]. Although effective, tissue clearing is an invasive and typically time-consuming procedure, which limits its applicability for preparation-free imaging of unaltered tissue slices, such as in fast point-of-care diagnostics.

Large-FOV imaging of thick, uncleared tissue slices, as well as imaging through diffusive and strongly scattering media, remains largely unexplored. In this contribution, we introduce Near-Infrared Lensless Digital Holographic Microscopy (NIR-LDHM), an approach that combines the complementary advantages of LDHM and NIR illumination while avoiding the optical complexity typically associated with NIR microscopy. In contrast to conventional NIR-QPI systems relying on specialized optics and expensive detectors, LDHM enables acquisition in both the VIS and NIR ranges using a standard, low-cost VIS CMOS detector. Although such sensors exhibit low quantum efficiency (QE) in the NIR, this does not limit the robustness of LDHM, as it performs well in the low-photon-budget (LPB) regime[26,27]. We implement this framework and provide a comprehensive comparative analysis of VIS and NIR performance across multiple controlled levels of scattering, identifying regimes in which VIS illumination fails while NIR remains effective. Our approach spans illumination wavelengths from the 480 nm VIS up to 1100 nm in the NIR - approaching the practical responsivity cutoff of silicon - and deliberately targets sample thicknesses for which multiple scattering becomes significant. Beyond wavelength selection, we identify an enabling novel mechanism in which increasing the sample–sensor propagation distance improves reconstruction quality under strong scattering, reframing distance – often simply minimized in lensless holography design - as a tunable acquisition parameter for robustness in turbid media. Finally, we extend the method to biological imaging of thick murine liver and brain tissue sections verifying high potential impact within biophotonics.

## 2    Materials and methods

The experimental system used throughout this study corresponds to the standard implementation of LDHM. It consists of a quasi-point source emitting spherical wave illumination, measured sample, and a board-level CMOS image sensor placed after the sample, with all elements arranged in an in-line configuration along a common optical axis. To systematically evaluate the system performance in a function of wavelength, six illumination wavelengths spanning VIS and NIR spectral ranges were employed throughout the study:

- 480 nm, 532 nm, 632 nm – generated using a supercontinuum laser source (NKT Photonics SuperK EVO) equipped with an acousto-optic tunable filter (SuperK SELECT) enabling wavelength selection; the filter provides spectral filtering with a full width at half maximum (FWHM) bandwidth of 1.8–8.5 nm. The source output provides a collimated beam that is delivered to the system through a 0.3 NA, 10× objective lens to produce a spherical wavefront,
- 785 nm - laser diode (Thorlabs LP785-SF20) pigtailed with a single-mode fiber with a 4.4 μm core diameter and 0.13 NA; FWHM = 3.0 nm,
- 978 nm - laser diode (Thorlabs LP980-SF15) pigtailed with a single-mode fiber with a 5.3 μm core diameter and 0.13 NA; FWHM = 1.7 nm,
- 1100 nm - generated using a supercontinuum laser source (NKT Photonics SuperK EVO) equipped with an acousto-optic tunable filter (SuperK SELECT) enabling wavelength selection; the filter provides spectral filtering with FWHM bandwidth of 6.4-19.8 nm. The source is coupled to a single-mode fiber with 8.8 ± 1.0 μm core diameter and 0.11 NA. The illumination wavelengths were selected to span the detection range of a regular CMOS sensor, extending from VIS to NIR (Alvium 1800 U-2050m monochrome bare board camera, Allied Vision; pixel size 2.4 μm × 2.4 μm, resolution 5496 × 3672), with spatial and temporal coherence of the same order of magnitude maintained across all sources.

The distance from the light source to the sample was fixed at approximately 300 mm for all experiments. At this separation, the illumination can be well approximated as a planar wave at the sample plane, while the ratio of the total system length (source-camera distance) to the source-sample distance results in an approximately constant magnification close to 1, as the overall system length is significantly larger than the sample-sensor distance. The image sensor was mounted on a linear translation stage (Thorlabs MTS25/M-Z8) that enabled controlled adjustment of the sensor distance relative to the sample. This capability was used for multi-plane hologram acquisition required for complex optical field recovery using an iterative multi-frame Gerchberg-Saxton (GS) reconstruction algorithm[28–30]. For each experimental condition, five images were recorded at different distances. The nearest measurement plane was positioned at 3.3 ± 0.5 mm from the sample, and each subsequent frame was acquired approximately $n \cdot 0.6$ mm farther from the sample (where $n = 1, …, 5$ is the number of the frame). These intensity measurements (holograms) were used as input to the multi-plane GS algorithm to reconstruct the complex optical field at the object plane.

To simulate scattering we employed well acknowledged scattering medium[31,32] - UHT pasteurized cow milk with 3.2% fat content. Whole milk provides a biologically relevant scattering model due to its lipid-droplet dominated structure and predominantly forward-scattering behavior, characterized by tissue-like reduced scattering coefficients[32]. In addition, whole milk exhibits short-term optical property stability for up to three hours[32], making it suitable for completing a full measurement set across all six analyzed wavelengths. The layer of scattering medium (scattering layer, SL), shown in Fig. 1, contains two square coverslips (each 170 μm thick) separated by a spacer, keeping the scattering medium inside such 'sandwich'. As a spacer we used 3M 467MP and 468MP double side acrylic adhesive tape, with nominal thickness of 50 μm and 130 μm, respectively. Such solution is favorable for customized-shape spacer implementation and high chemical resistance. To achieve thickness higher than nominal 50 um and 130 um, we performed stacking of tape layers.

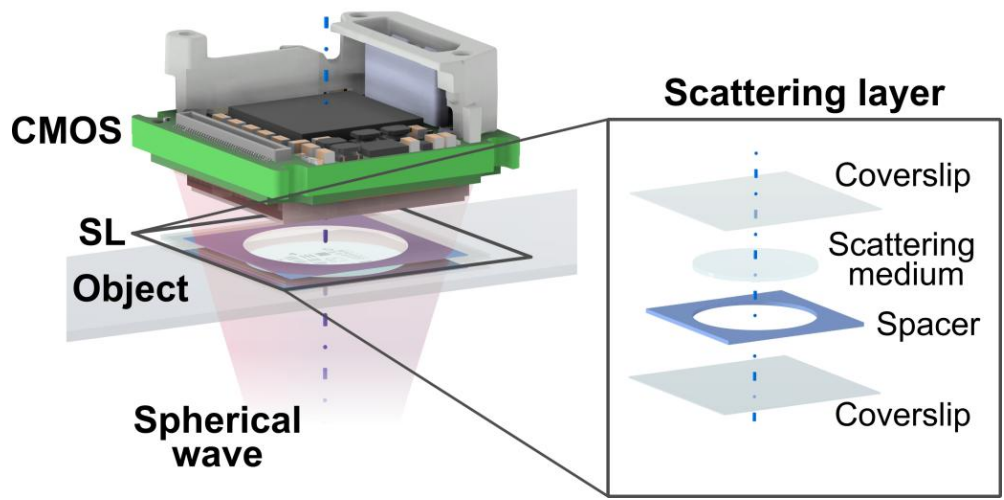

**Fig. 1** LDHM system configured for controlled scattering experiments.

Under realistic biological imaging conditions, the scattering medium is an integral part of the specimen: it surrounds the structures of interest and remains in direct physical contact with them. To faithfully emulate such conditions in a controlled simulated experiment, the test target should therefore be immersed within the scattering medium. This configuration most closely reproduces *in situ* biological scattering. However, direct contact between milk-based scattering medium and high-quality resolution targets is undesirable, as it may contaminate or degrade the targets and compromise measurement reliability. Such constraints are typical in experimental practice and often impose the use of separated scattering medium[33–35]. The simplicity of LDHM makes it particularly well suited to systematically investigate and compare alternative experimental configurations in which the milk-based SL is spatially separated from the test target while preserving controlled scattering conditions. The detailed analysis of several system configurations was performed and further covered in the Supplementary Material, Supplementary Note 1. These results show that the configuration in which the SL is placed between the sample and the detector exhibits resolution performance nearly identical to that of the fully immersed reference configuration. Accordingly, this configuration - Fig. 1 - is adopted throughout the study.

## 3 Results

**Investigating NIR-LDHM phase and amplitude imaging under high- and low-photon-budget regimes**

One of remarkable advantages of LDHM is its ability to operate at extremely low illumination doses[26,27,36]. This capability is favorably exploited to minimize photostimulation[37] and phototoxicity[38] when imaging living biological specimens and is referred to here as the LPB regime. In the context of NIR-LDHM, this feature becomes particularly beneficial, as standard VIS-optimized board-level CMOS detectors exhibit only a residual QE in the NIR illumination conditions. Despite the sharply reduced detector responsiveness at longer wavelengths - decreasing to 0.19% QE at 1100 nm, see Fig. 2(a) - the CMOS sensor used in this study is nevertheless capable of recording holograms with sufficient signal-to-noise ratio (SNR) to maintain system resolution. Here, the HPB regime is defined as the signal exceeding 100 gray levels at recorded at the detector, whereas LPB corresponds to signal in the range of approximately 5–15 gray values.

To ensure a consistent comparison across wavelengths, the optical power incident at the sample plane was kept constant at 0.35 µW for all measurements; average power density down to ultralow 4.9 nW/mm² over the sensor aperture (Thorlabs optical power meter S121C sensor, 9.5 mm aperture diameter). Exposure times were then adjusted to reach either HPB or LPB conditions. HPB reference measurements were first established in the VIS range (exp = 0.11 s), and the corresponding exposure time was subsequently applied to NIR acquisitions. Under these conditions, the 785 nm illumination naturally fell within the LPB regime, whereas longer NIR wavelengths required moderate increases (around 0.30 s and 0.55 s) in exposure time to reach LPB signal levels due to reduced detector QE. For 978 nm and 1100 nm, this resulted in slightly lower reconstructed resolution, which we attribute primarily to the limited SNR and the use of a basic GS reconstruction; advanced filtering approaches[26,27] are expected to mitigate this effect. Increasing the exposure time further enabled HPB operation for all NIR wavelengths (0.30 s for both tests in 785 nm, 1.11 s for amplitude and 1.81 s for phase tests in 978 nm and 7.14 s for both tests in 1100 nm). In this regime, only minor resolution differences were observed relative to VIS, which theoretical LDHM analysis indicates arise mainly from variations in illumination coherence rather than wavelength increase.

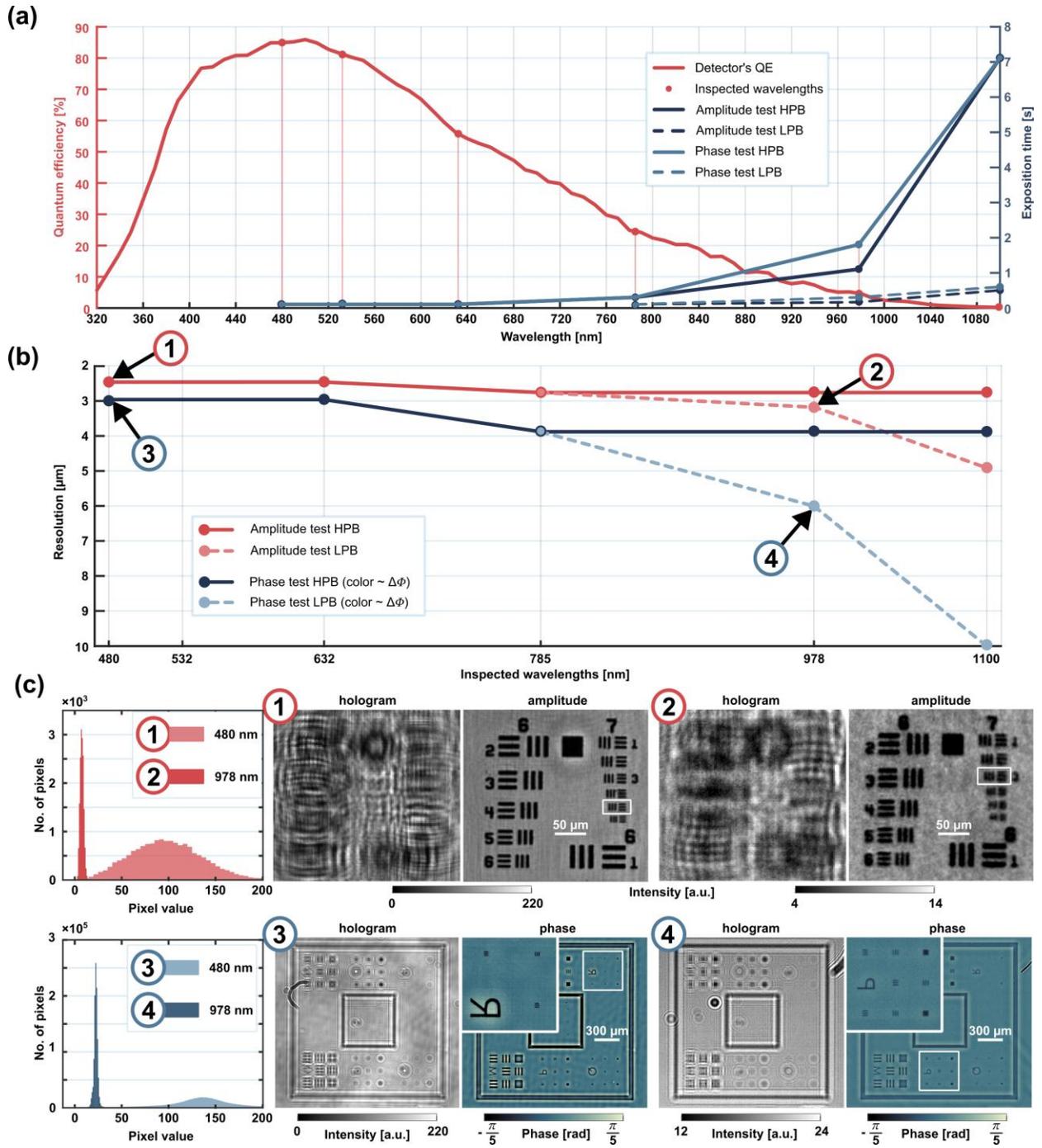

**Fig. 2** LPB operation of LDHM in the VIS and NIR. (a) QE of the CMOS detector as a function of wavelength, together with the exposure times required for HPB and LPB acquisition conditions. (b) Best achievable lateral resolution for amplitude and phase targets as a function of illumination wavelength under HPB (solid lines) and LPB (dotted lines) conditions. (c) Reconstructed amplitude (1-2) and phase (3-4) maps from the respective holograms acquired at 480 nm (1, 3) under HPB and at 978 nm (2, 4), under LPB conditions, along with representative histograms. White boxes indicate the smallest resolvable features.

It is important to note that the phase resolution target used in these measurements was a Borofloat 33 glass substrate etched to a depth of 125 nm. The corresponding phase delay is inherently wavelength dependent, decreasing with increasing illumination wavelength. Consequently, as the illumination wavelength shifts toward the NIR, the phase modulation imposed by the target becomes weaker, leading to a reduced SNR. Figure 2(b) summarizes the smallest resolvable features obtained under different measurement conditions. The solid lines in Fig. 2(b) correspond to HPB measurements across all six wavelengths, where HPB means acquisition conditions in which the recorded intensity histograms are uniformly distributed across the full dynamic range of the detector, with exposure time adjusted accordingly.

Under the LDHM configuration employed here, and for the distances between elements described in Section 2, the theoretical lateral resolution is limited by the detector pixel size (2.4 × 2.4 µm$^2$), rather than by the illumination wavelength within the analyzed range (480-1100 nm). Nevertheless, a slight degradation in resolution is observed toward longer wavelengths in Fig. 2(b). This effect is attributed probably due to residual differences in the spatiotemporal coherence properties of the illumination sources[39]. Although coherence conditions were matched as closely as possible, remaining variations introduce modest wavelength-dependent performance differences. As a result, subsequent analyses in this study emphasize relative changes with respect to the non-scattering reference case.

The dotted curves in Fig. 2(b) correspond to LPB measurements at NIR wavelengths, acquired with a fixed exposure time of 0.11 s, corresponding to the optimal HPB exposure for VIS illumination. Under these constant-exposure conditions, a noticeable reduction in resolution is observed at 978 nm and 1100 nm for both amplitude and phase imaging. This effect is further illustrated in Fig. 2(c), which compares intensity histograms for amplitude and phase holograms acquired under HPB conditions at 480 nm and LPB conditions at 978 nm, using the constant exposure time of 0.11 s. While the 480 nm measurements utilize approximately 200 effective intensity levels to encode the holograms, the 978 nm measurements rely on only 10 intensity levels, resulting in a modest resolution reduction despite the extremely limited photon budget.

**Wavelength-dependent LDHM phase and amplitude resolution analysis under increasing scattering**

For the initial phase-resolution measurements, the smallest feature size on the phase target was 2.0 µm, which is both theoretically and experimentally unresolved under the present system constraints. Consequently, the next available feature size of 2.96 µm defines the effective lateral resolution limit for phase imaging in Fig. 2. For the amplitude measurements, performed using a standard positive USAF resolution target (groups 2-7, Thorlabs), the smallest theoretically resolvable bar width was 2.46 µm. To enable more sensitive detection of subtle resolution changes under scattering conditions, the phase target was subsequently replaced with a quantitative phase USAF target (Benchmark Technologies) comprising groups 6-10. Here, we studied 100 nm-high structures with a RI of approximately 1.52.

The amplitude and phase USAF targets were measured under progressively increasing scattering conditions, realized by introducing SLs of controlled thickness. Measurements began with no SL. Figure 3(a) summarizes the best achievable lateral resolution as a function of SL thickness for both amplitude (solid lines) and phase imaging (dotted lines) across the six investigated illumination wavelengths, yielding a total of twelve resolution curves. Although the scattering thickness was varied discretely, the measured data points are connected by lines in Fig. 3(a) to facilitate visualization of overall trends across the large parameter space. Any potential uncertainties in manually reading the resolution do not affect the observed trends. For these measurements, exposure times were adjusted individually to achieve HPB measurement, accounting for differences in detector's quantum efficiency for each illumination wavelength and SL thickness. The illumination power at the sample plane was kept constant for each wavelength at approximately 67 µW, corresponding to an average optical power density of 950 nW/mm$^2$. For completeness, reconstructed amplitude and phase maps for selected SL thicknesses are provided in Supplementary Material, Supplementary Note 2 for direct visual reference.

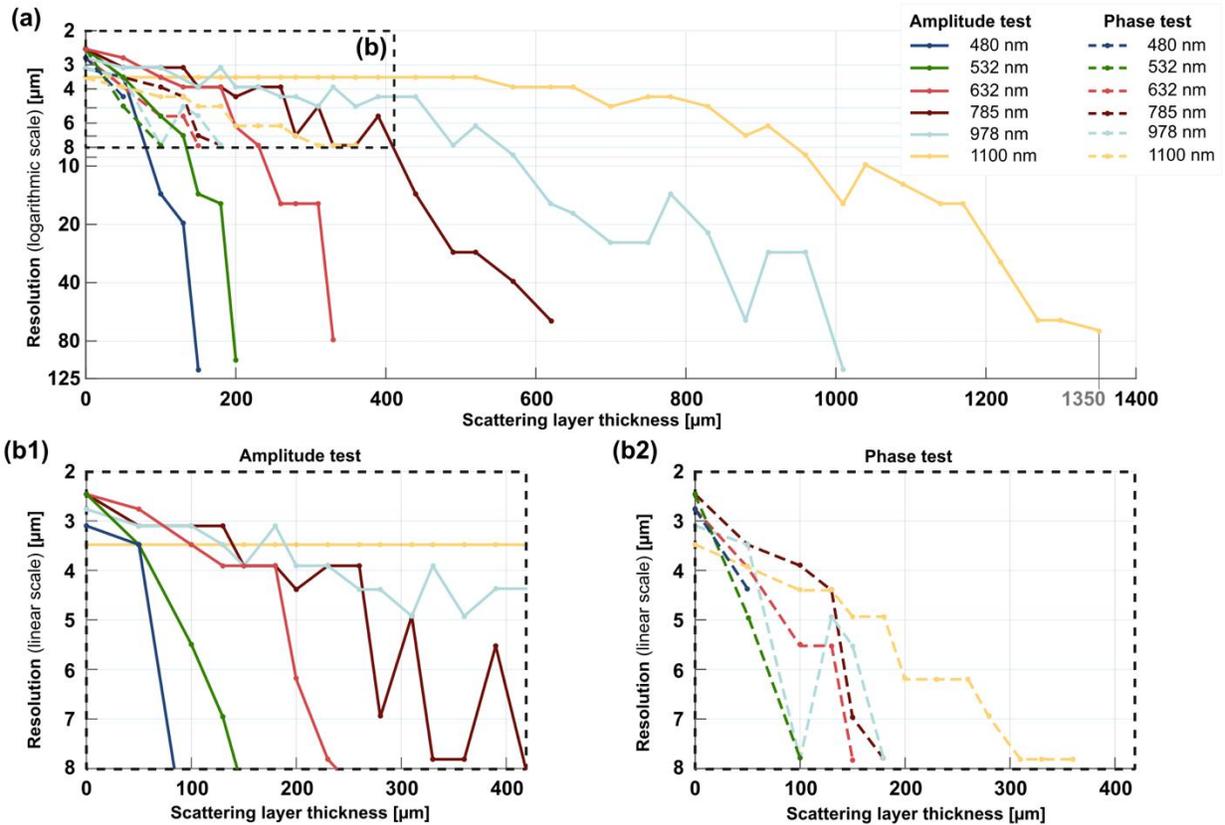

**Fig. 3** Wavelength-dependent LDHM resolution under increasing scattering. (a) Lateral resolution for amplitude (solid lines) and phase (dotted lines) tests imaging versus SL thickness increase at six illumination wavelengths, shown on a logarithmic scale. The highlighted region is expanded in panels (b1) and (b2). (b1) Amplitude-only resolution plotted on a linear scale over the same resolution range as the phase data. (b2) Phase-only resolution plotted on a linear scale, limited by the USAF feature sizes available on the quantitative phase target.

The results in Fig. 3 demonstrate a performance advantage of NIR-LDHM compared to conventional VIS-wavelengths operation. Overall, the data reveal a clear trend of increasing robustness to strong scattering with increasing illumination wavelength. Importantly, the magnitude of this improvement is substantial. While VIS illumination becomes fully unresolved at SL thicknesses of 180 μm, 230 μm, and 360 μm for wavelengths of 480 nm, 532 nm, and 632 nm, respectively, NIR illumination maintains resolvable features up to critically greater thicknesses of approximately 650 μm, 1040 μm, and 1400 μm for 785 nm, 978 nm, and 1100 nm, respectively. These results demonstrate a crucial extension of the operational regime of LDHM into multiple scattering conditions, well beyond what is to date achievable with VIS illumination. Moreover, for amplitude imaging with 1100 nm wavelength, no measurable degradation in lateral resolution is observed for SL thicknesses up to 570 µm, further highlighting the robustness of the NIR-LDHM approach in multiple scattering media.

Comparison between amplitude and phase resolution trends reveals systematically higher sensitivity of phase signal to multiple scattering, although qualitatively it follows a similar trend to that observed for amplitude signal, however, over a narrower range of SL thicknesses. This behavior is primarily a consequence of the wavelength dependence of the phase contrast: as the illumination wavelength increases, the phase shift induced by a fixed-height structure decreases, resulting in a reduced phase SNR. In contrast, the amplitude signal exhibits comparable intrinsic SNR across wavelengths, with SNR degradation arising predominantly from increased scattering rather than wavelength-dependent contrast loss.

Since the quantitative phase target employed in this part of the study comprised only USAF groups 6-10, phase resolution measurements were limited to feature sizes down to 7.81 µm (element 1 of group 6). Notably, despite the increased sensitivity of phase imaging to scattering, phase measurements at 1100 nm retain resolvable features at substantially greater SL thicknesses than amplitude measurements at shorter wavelengths. Specifically, at a SL thickness of 360 µm, phase imaging at 1100 nm resolves features down to 7.81 µm, whereas amplitude imaging at 632 nm degrades to a resolution of approximately 78.75 µm at SL thickness of 330 µm. This comparison highlights the significant advantage conferred by NIR illumination for phase imaging in strongly scattering media.

**A new mechanism for increasing LDHM capabilities by adjusting sample-sensor distance**

Beyond wavelength effects, the sample-sensor propagation distance plays a critical role in LDHM performance under multiple scattering. In the weakly-scattering regime, where the source-sample distance is much larger than the sample-sensor distance, reducing the sample-sensor separation generally improves the effective system's NA and reconstruction quality, as observed in previous LDHM implementations[40,41]. Accordingly, for standard measurement, the five holograms used as an input to the multi-height GS reconstruction algorithm were acquired over a narrow sample-sensor distance of approximately 3-6 mm, with a 0.6 mm separation between each hologram. While measurements without a SL could be performed at distances down to ~1.5 mm, introducing the milk-based SL required additional space; therefore, 3 mm was adopted as the minimal sample-sensor distance.

During reconstruction of the results shown in Fig. 3 an interesting trend was observed. Above a certain scattering level, the conventional LDHM intuition - that shorter sample-sensor distances yield superior resolution[39] - no longer held. Instead, improved resolution was obtained at larger propagation distances. Notably, this improvement occurred despite a substantial decrease in the detected intensity at larger distances. To investigate this effect systematically, a specialized study was performed at an illumination wavelength of 632 nm. A total of 52 holograms of the amplitude USAF target were acquired, starting from an initial sample-sensor distance of approximately 3 mm up to around 45 mm. First 35 frames were recorded at increments of 0.3 mm and another 7 with 1 mm, and the rest frames with 2 mm separation.

In standard measurements using five frames, no exposure-time adjustment was required across the acquisition planes; therefore, the same acquisition conditions were adopted for this extended dataset. The exposure time was set to optimally record the closest hologram and then kept constant for all subsequent frames, resulting in progressively lower intensity level for holograms recorded at larger sample-camera distances. Each frame was reconstructed individually using the angular spectrum (AS) backpropagation algorithm[42], which also constitutes the propagation step within the multi-frame GS reconstruction used for data reconstructed in previous subsections. Despite the reduced intensity, improved resolution was consistently observed for the more distant holograms under strong scattering conditions. Measurements were conducted for three SL's thicknesses (Fig. 4(a1)): 100 μm (weak scattering), 200 μm (moderate scattering), and 300 μm (strong scattering). Figure 3 shows that amplitude features become unresolvable for SL thicknesses above ~330 μm at 632 nm, therefore, the 300 μm case lies close to the resolution limit.

Figure 4(b) summarizes the results obtained by independently reconstructing each hologram as a function of propagation distance. Under weak scattering conditions (100 μm SL), resolution improves slightly as the sample-sensor distance increases, reaching approximately constant value over the range of ~5-11 mm, and then undergoes a gradual degradation at larger distances (≈12 mm and beyond). In contrast, under moderate and strong scattering conditions (200 μm and 300 μm SLs), the behavior changes. In these regimes, the reconstructed resolution improves with increasing propagation distance up to approximately 14 mm, followed by a slow decline with distances in the range of ~15-45 mm. At the same time, the spread of the measured resolution values increases with scattering strength, consistent with a reduced SNR at higher scattering levels. At sample-sensor distances approaching ~40 mm, the resolution in the strong scattering regime degrades back to values comparable to those obtained at the smallest distances. Notably, in the strong scattering regime, increasing the sample-sensor distance by a factor of four - from approximately 3 mm to 12 mm - results in an approximately twofold improvement in resolution, from ~11 μm to ~5.5 μm. This interesting mechanistic behavior highlights the nontrivial role of propagation distance in LDHM under strongly scattering conditions. For visual reference representative examples of the nine holograms, selected across ~13 mm distance range, and their individually reconstructed amplitude maps, for the highest scattering case, are shown in Fig. S8 in Supplementary Material, Supplementary Note 3.

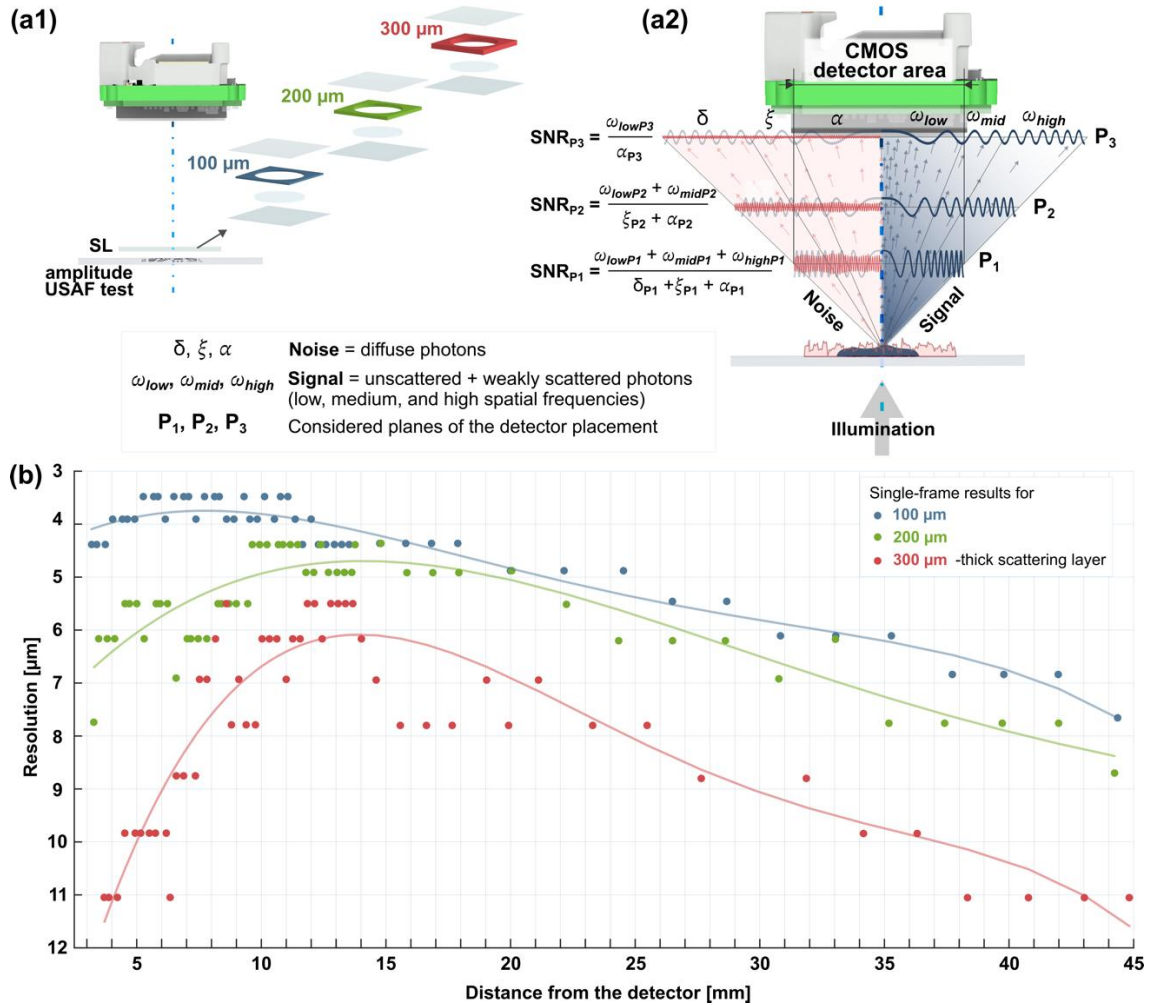

**Fig. 4** Sample-detector-dependent resolution under scattering. (a1) Schematic illustration of photon contributions and underlying mechanisms at detector planes $P_1$, $P_2$, $P_3$ with increasing sample-sensor distances. (a2) Experimental configuration used for the propagation-distance study. (b) Resolution of each individual frame in a function sample-sensor distance at 632 nm for three scattering levels. A fourth-degree polynomial was fitted to the discrete data points to visualize the overall trend.

The observed improvement in reconstruction quality with increasing sample-sensor distance can be understood by considering the angular and spatial-frequency composition of the detected field (Fig. 4(a2)). In LDHM, the recorded hologram arises from the interference of the unscattered incident illumination, that in our system can be approximated as plane wave, and weakly scattered wave generated by the object. In such conditions the resulting signal is confined, in a statistical sense, to a narrower angular cone than the diffusely scattered background. Consequently, signal-carrying photons occupy smaller effective scattering angle than noise photons. In the detector plane, Gabor fringes encode spatial-frequency information of the object. For a single point at the optical axis, lower spatial frequencies are encoded closer to the axis, while higher spatial frequencies appear at larger radial distances from the axis. The finest fringes, corresponding to the highest spatial frequencies ($\omega_{high}$) of the object, are therefore located farthest from the optical axis and exhibit the lowest contrast. Although the overall signal amplitude is highest at shorter propagation distances, plane $P_1$, it decreases with increasing distance due to geometric spreading; however, this decrease is slower than that of the noise component.

Diffuse photons, which constitute noise, arise from highly scattered light, and most diffuse photons are distributed over a much wider angular range. Their contribution to the detected field is broadband in spatial frequency and manifests as pixel-to-pixel intensity fluctuations largely independent of propagation distance ($P_1$ - $P_3$). Because of their wide angular distribution, the effective cone containing most diffuse photons expands more rapidly with distance than that of the signal, causing the amplitude of the detected noise to decrease faster than that of the signal as the sample-detector distance increases. As a result, at the closest detector plane ($P_1$), the system collects most signal components, including low, medium, and high ($\omega_{lowP1}$, $\omega_{midP1}$, $\omega_{highP1}$) spatial frequencies, but also a majority of the noise ($\delta_{P1}$, $\xi_{P1}$, $\alpha_{P1}$). Under these conditions, the noise overwhelms the weaker $\omega_{midP1}$ and $\omega_{highP1}$ signal components, and only low spatial frequencies $\omega_{lowP1}$ are effectively resolved. At an intermediate plane ($P_2$), the effective NA is reduced, and the highest spatial frequencies $\omega_{highP2}$ and $\delta_{P2}$ part of the noise are no longer collected. However, because the noise decreases more rapidly with distance than the signal, the SNR improves, enabling detection of both $\omega_{lowP2}$ and $\omega_{midP2}$ components. At a larger propagation distance ($P_3$), the geometrically collected signal is limited primarily to $\omega_{lowP3}$ spatial frequencies, but the diffuse background $\alpha_{P3}$ is strongly suppressed, and the remaining signal exhibits relatively high contrast due to the substantially reduced noise contribution.

In other words, increasing the sample-camera distance constitutes a mechanism which filters out the noise and augments the SNR of the hologram. It improves the resolution due to generally low NA of LDHM, which is pixel limited and not diffraction limited. When the distance becomes too big the low NA signal starts to be filtered out as well, lowering the effective NA below the pixel limitation. Thus, a further gradual decrease of the resolution is observed.

**Biological verification for *in vitro* NIR-LDHM imaging**

We next apply the insights gained from controlled scattering experiments to biologically relevant specimens. Thick tissue slices - specifically mouse liver (Fig. 5) and mouse brain (Fig. 6) - present strongly scattering and heterogeneous conditions under which conventional label-free phase imaging is known to degrade, especially these operating in Gabor regime. Using the NIR-LDHM approach, we demonstrate large FOV amplitude and phase imaging of these tissues under challenging optical conditions. For both tissue types, we directly compare VIS (532 nm) and NIR (1100 nm) illumination. Each figure presents reconstructed amplitude and phase maps, with representative regions of interest enlarged to highlight internal structural features. Regions of interest might differ between amplitude and phase maps, since relevant morphological features might be visible in one or the other modality. In Fig. 5, liver slices with thicknesses of 20, 30, and 60 μm were examined, while Fig. 6 shows brain slices with thicknesses of 50, 125, 150, 200, and 250 μm. Details of sample preparation and enlarged results of the brain slices are provided in Supplementary Material, Supplementary Note 4. Differences in the achievable imaging depth between the two tissue types reflect their intrinsic scattering properties. Liver tissue, which is optically denser and more strongly scattering than brain tissue, reaches the practical imaging limit up to 60 μm even under NIR illumination. In comparison, brain tissue remains distinguishable to NIR-LDHM imaging at thicknesses up to approximately 250 μm, highlighting the tissue-dependent nature of scattering and attenuation in biological samples.

Across all examined thicknesses, illumination at 1100 nm consistently reveals finer internal features (higher resolution) compared to 532 nm illumination, where such details are largely obscured by noise. Under VIS illumination, both amplitude and phase reconstructions predominantly capture just the shape of the tissue, with limited contrast arising from internal microstructure. In contrast, NIR imaging preserves meaningful internal contrast in both amplitude and phase, enabling visualization of tissue organization beyond gross morphology.

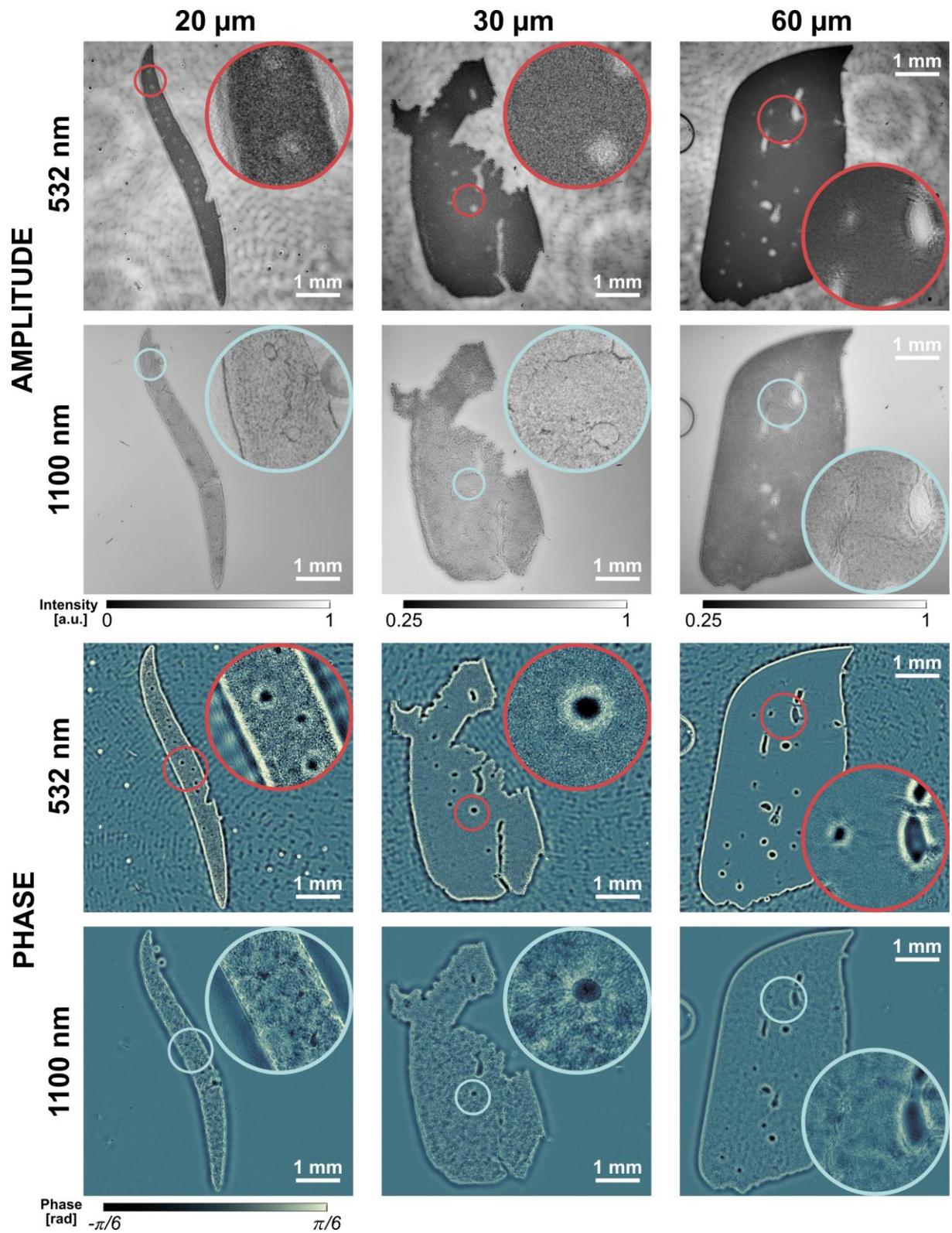

**Fig. 5** Mouse liver imaging with VIS- and NIR-LDHM. Amplitude and phase maps acquired at 532 nm and 1100 nm for liver slices of increasing thickness, showing enhanced internal contrast under NIR illumination, revealing trabecular patterns and vascular pathways.

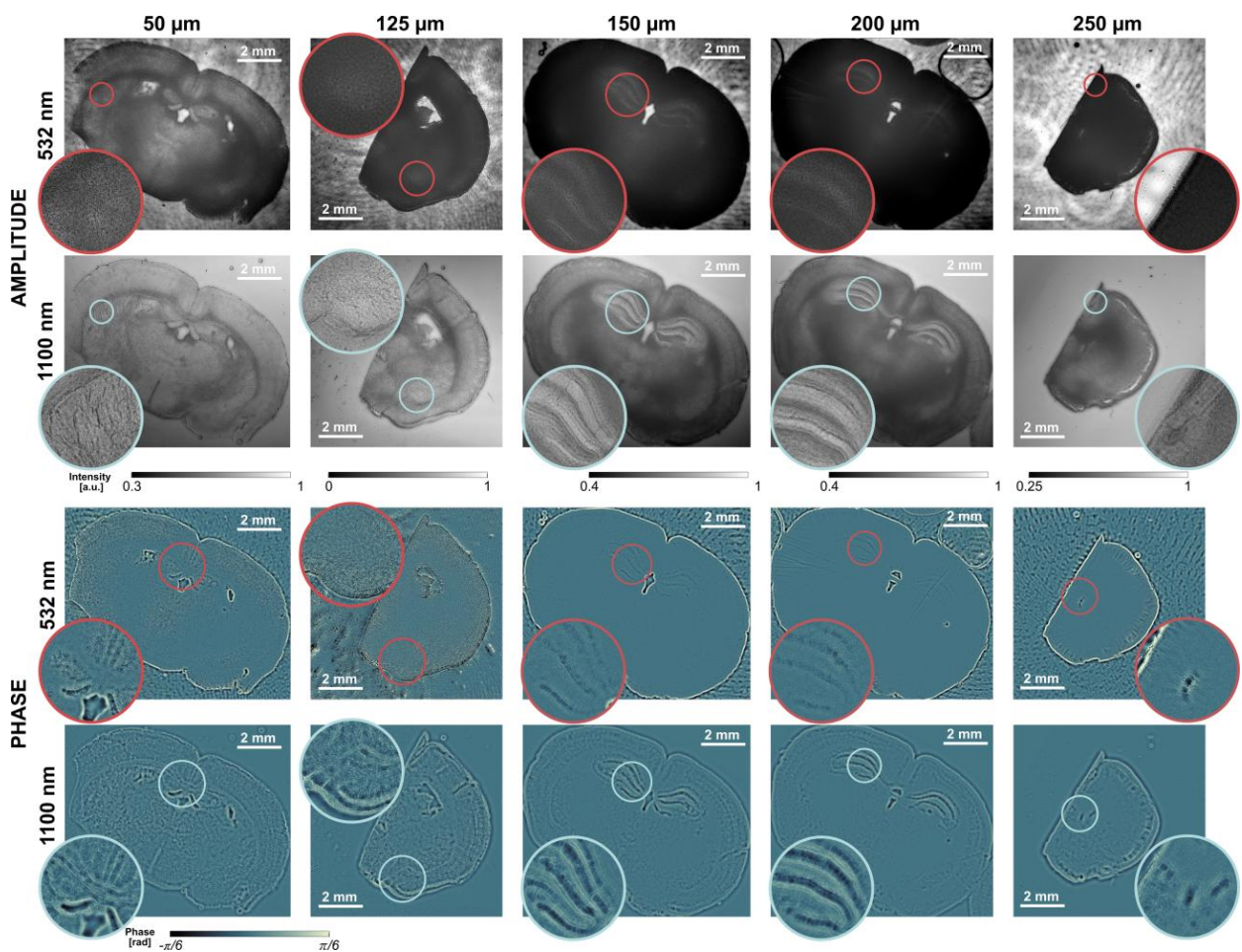

**Fig. 6** Brain tissue imaging using VIS- and NIR-LDHM. Amplitude and phase maps acquired at 532 nm and 1100 nm for brain slices of increasing thickness. Enlarged images of brain tissue VIS- and NIR-LDHM results are presented in Figs. S10-S14 in the Supplementary Note 4.

For mouse liver, NIR illumination consistently outperforms VIS illumination across all investigated slice thicknesses, with the magnitude of improvement increasing with depth. In thin sections (20 and 30 µm), NIR imaging already provides higher contrast and more continuous delineation of internal tissue features compared with VIS illumination, facilitating clearer visualization of trabecular organization and vascular structures. While VIS imaging remains partially informative at these thicknesses, internal features appear less distinct and more susceptible to scattering-induced noise. The advantage of NIR illumination becomes pronounced in thicker liver sections (60 µm). At this thickness, VIS illumination fails to preserve meaningful internal contrast, and reconstructions are largely limited to overall tissue outlines. In contrast, NIR imaging at 1100 nm retains discernible internal architecture, including trabecular patterns and vascular pathways, enabling more reliable morphological interpretation. Due to the intrinsically high scattering and optical density of liver tissue, the practical imaging depth of NIR-LDHM is reached at approximately 60 µm.

Mouse brain slices exhibit a laminar organization with spatially varying fiber density, providing a sensitive test for depth-resolved imaging performance. While both VIS and NIR illumination reveal aspects of tissue organization in thin sections, the advantage of NIR illumination becomes pronounced with increasing thickness. For brain slices thicker than 150 µm, NIR imaging at 1100 nm consistently yields higher contrast and clearer visualization of layered structures than VIS illumination. Internal spatial relationships remain distinguishable in NIR images up to thicknesses of approximately 250 µm. In contrast, VIS imaging suffers from strong contrast loss in thicker sections, and increasing signal intensity improves only local contrast without restoring clear structural interpretation. Overall, brain tissue remains susceptible to NIR-LDHM imaging at substantially greater thickness than liver tissue, underscoring both the reduced scattering of brain tissue and the superior penetration capability of NIR illumination. Importantly, this advantage is achieved without the need for optical clearing, which is often required for VIS imaging of thick tissues.

## 4  Discussion

Although major quantitative conclusions in this study rely on amplitude targets to isolate scattering-induced degradation, we also demonstrate that NIR-LDHM supports phase imaging under the same conditions: qualitative phase results presented in Fig. S15 in Supplementary Material, Supplementary Note 5. Phase contrast from a fixed-height structure decreases with increasing wavelength because the phase delay scales as $\Delta\varphi = (2\pi/\lambda)\, \Delta OPD$. Consequently, at longer wavelengths the phase SNR becomes increasingly contrast-limited and noise-limited, assuming the scattering medium is not purely phase and does not undergo the same decrease effect as phase object. This explains the greater sensitivity to scattering observed for phase compared with amplitude imaging at a given wavelength. Nevertheless, our results show that phase NIR-LDHM imaging remains practically useful even when the phase modulation is substantially reduced. This highlights a key practical advantage of NIR-LDHM: despite operating with a VIS-optimized detector and, in some cases, under LPB conditions, the system can recover subtle phase variations at 1100 nm. Moreover, in specimens that introduce large phase delays - such as thick or high-refractive-index structures - NIR illumination provides a straightforward means to reduce phase delay and unwanted phase discontinuities without resorting to index-matching media. This flexibility is particularly attractive in LDHM systems, where there is no penalty from absorption and reflectance losses that typically limit performance in conventional lens-based microscopy at comparable wavelengths. Importantly, this configuration enables a direct comparison of scattering effects across the VIS-NIR range without any modification of the optical system apart from the illumination wavelength, allowing the observed differences to be attributed primarily to wavelength-dependent propagation and scattering.

Our systematic distance study shows that, at high scattering levels, increasing the sample-sensor distance can improve effective resolution - even when the recorded hologram becomes substantially dimmer and the exposure time is held constant. We report this effect as a consistent trend observed across our measurements; however, we emphasize that the relationship between scattering and propagation distance requires more thorough investigation. The present results provide an initial indication of distance-dependent scattering behavior, but a more detailed study is needed to fully understand the underlying mechanisms. Further work addressing scattering as a function of distance from the sample is currently ongoing and will be reported in future studies.

The biological experiments confirm that the insights obtained from controlled scattering translate to real tissues. These results, however, emphasize that imaging performance is governed not simply by tissue thickness but by tissue-specific scattering and attenuation properties. It should be noted that while SLs provide controlled and reproducible testing conditions, they represent only one type of scattering medium and do not fully reflect the diversity of real biological environments. In actual tissues, scattering is inherently more complex and we acknowledge that using milk phantoms does not fully reflect the heterogeneous nature of biological samples.

An additional practical implication of this study is that NIR lensless holography up to 1100 nm is feasible on standard silicon hardware, but achievable performance is inherently detector-dependent. Not all VIS-optimized CMOS sensors exhibit a sufficiently long responsivity tail into the NIR; in many cases, the effective cutoff near the VIS-NIR boundary is determined not only by the silicon itself but also by integrated detector elements such as IR-blocking filters or protective cover glass. These components can significantly attenuate NIR radiation depending on their thickness and spectral characteristics. In our system, these elements did not fully block NIR radiation, enabling operation up to 1100 nm; however, this may not be the case for all VIS-optimized sensors. Our results demonstrate that even residual QE can be sufficient for robust hologram acquisition and reconstruction, because LDHM can operate effectively in LPB regimes where many other imaging modalities fail. While the use of specialized NIR detectors would likely further improve performance by reducing exposure-time requirements and increasing SNR, employing a single off-the-shelf VIS detector across both VIS and NIR wavelengths offers substantial advantages in cost, system simplicity, and experimental flexibility. This is particularly important in LDHM, where conventional VIS cameras typically offer much smaller pixel sizes - enabling higher spatial resolution - and higher pixel counts - supporting a larger FOV - than short-wave infrared (SWIR) detectors. The primary trade-off of operating near the silicon cutoff is the need for longer exposure times, which may limit imaging of rapidly dynamic specimens; however, this limitation can be mitigated through increased illumination power or more sensitive detector technologies. These considerations point toward clear hardware upgrade paths that preserve the core simplicity of the LDHM architecture while further extending its performance in strongly scattering samples.

Several practical challenges remain. Wavelength-specific parasitic fringes are observed near 800-1000 nm for our detector and can introduce additional variability (further described in Fig. S16 in the Supplementary Material, Supplementary Note 6). These artifacts may be linked to mismatches in spectral and temporal coherences across the light sources used, highlighting how sensitive LDHM measurements in scattering conditions are to coherence properties. This also underscores the importance of coherence control, which is crucial for LDHM imaging in scattering conditions yet difficult to perfectly match across a broad wavelength range. While these artifacts do not alter the overall trends, mitigation strategies based on improved spectral and temporal coherence control could further enhance robustness. Despite these constraints, the simplicity of the architecture - single board-level sensor, transmission geometry, and absence of imaging optics - provides a practical framework for further exploration of deep-tissue quantitative phase imaging and imaging in demanding scattering conditions.

## 5  Conclusion

This work establishes near-infrared lensless digital holographic microscopy (NIR-LDHM) as a practical route to extend in-line Gabor holography from the traditional weak-scattering regime toward optically thick, multiply scattering specimens - without abandoning the simplicity, compact footprint, and low-cost that make lensless architectures attractive. By shifting illumination from the visible into the near-infrared, up to the silicon cutoff (~1100 nm), we strengthen the physical conditions required for high-fidelity hologram formation: a larger fraction of the object field remains detectable after propagation through turbid layers, the unscattered reference wave is less depleted, and quantitative recovery of both amplitude and phase remains feasible in regimes where visible-wavelength operation fails. Under controlled tissue-mimicking scattering conditions, we map reconstruction performance versus wavelength, scattering thickness, and sample-sensor distance, demonstrating that NIR operation preserves resolvable spatial frequencies through millimeter-scale scattering layers (up to ~1.4 mm at 1100 nm), far beyond the breakdown observed in the visible. Importantly, we show that these gains remain accessible under low-photon-budget acquisition using a standard board-level CMOS sensor, indicating that improved performance does not require specialized NIR detectors, a critical feature which increases low-cost accessibility.

Beyond wavelength selection, we identify an enabling mechanism in which increasing the sample-sensor propagation distance improves reconstruction quality under strong scattering, which we attribute to improved detectability of the interference pattern at the detector plane in this regime. This observation reframes distance - often minimized in lensless holography - as a tunable acquisition parameter for robustness in turbid media and motivates distance-aware acquisition strategies and scattering-adapted reconstruction algorithms.

Finally, we translate these advances to biologically relevant imaging by demonstrating large-field-of-view, label-free amplitude and phase imaging of uncleared *in vitro* mouse liver and brain slices, where 1100 nm illumination consistently yields higher internal contrast and more interpretable amplitude/phase structure than visible illumination across increasing thickness (up to ~250 µm in brain). Together, these results position NIR-LDHM as a scalable, hardware-minimal approach for high-throughput complex-field imaging through scattering media with immediate relevance to computational holography, imaging through complex media and biophotonics. Future extensions - such as scattering-aware inverse models with learned or physics-informed priors and detectors beyond silicon - provide clear pathways to further improvements in penetration depth, robustness, and acquisition speed, and to deployment in compact and resource-limited settings.

*Acknowledgments*

This work has been funded by the National Centre for Research and Development, Poland (LIDER14/0329/2023 in Lider XIV call). The research was carried out on devices co-founded by the Warsaw University of Technology within the Excellence Initiative: Research University (IDUB) programme.

*References*

1. G. Popescu, *Quantitative Phase Imaging of Cells and Tissues*, McGraw-Hill, New York (2023).
2. Y. Park, C. Depeursinge, and G. Popescu, "Quantitative phase imaging in biomedicine," Nature Photon **12**(10), 578–589 (2018) [doi:10.1038/s41566-018-0253-x].
3. T. L. Nguyen et al., "Quantitative Phase Imaging: Recent Advances and Expanding Potential in Biomedicine," ACS Nano **16**(8), 11516–11544, American Chemical Society (2022) [doi:10.1021/acsnano.1c11507].
4. P. Marquet, C. Depeursinge, and P. J. Magistretti, "Review of quantitative phase-digital holographic microscopy: promising novel imaging technique to resolve neuronal network activity and identify cellular biomarkers of psychiatric disorders," NPh **1**(2), 020901, SPIE (2014) [doi:10.1117/1.NPh.1.2.020901].
5. D. Gabor, "A New Microscopic Principle | Nature," 777–778 (1948) [doi:https://doi.org/10.1038/161777a0].
6. J. Garcia-Sucerquia et al., "Digital in-line holographic microscopy," Applied Optics **45**(5), 836 (2006) [doi:https://doi.org/10.1364/AO.45.000836].
7. Y. Wu and A. Ozcan, "Lensless digital holographic microscopy and its applications in biomedicine and environmental monitoring," Methods **136**, 4–16 (2018) [doi:10.1016/j.ymeth.2017.08.013].
8. A. Greenbaum et al., "Imaging without lenses: achievements and remaining challenges of wide-field on-chip microscopy," Nat Methods **9**(9), 889–895 (2012) [doi:10.1038/nmeth.2114].
9. G. Kim et al., "Holotomography," Nat Rev Methods Primers **4**(1), 51, Nature Publishing Group (2024) [doi:10.1038/s43586-024-00327-1].
10. H. Hugonnet et al., "Multiscale label-free volumetric holographic histopathology of thick-tissue slides with subcellular resolution," AP **3**(2), 026004, SPIE (2021) [doi:10.1117/1.AP.3.2.026004].
11. T. Li et al., "Reflection-mode diffraction tomography of multiple-scattering samples on a reflective substrate from intensity images," Optica, OPTICA **12**(3), 406–417, Optica Publishing Group (2025) [doi:10.1364/OPTICA.547372].
12. M. Chen et al., "Multi-layer Born multiple-scattering model for 3D phase microscopy," Optica, OPTICA **7**(5), 394–403, Optica Publishing Group (2020) [doi:10.1364/OPTICA.383030].

13. M. Lee, H. Hugonnet, and Y. Park, "Inverse problem solver for multiple light scattering using modified Born series," Optica, OPTICA **9**(2), 177–182, Optica Publishing Group (2022) [doi:10.1364/OPTICA.446511].
14. S. Chowdhury et al., "High-resolution 3D refractive index microscopy of multiple-scattering samples from intensity images," Optica, OPTICA **6**(9), 1211–1219, Optica Publishing Group (2019) [doi:10.1364/OPTICA.6.001211].
15. O. Yasuhiko and K. Takeuchi, "In-silico clearing approach for deep refractive index tomography by partial reconstruction and wave-backpropagation," Light Sci Appl **12**(1), 101, Nature Publishing Group (2023) [doi:10.1038/s41377-023-01144-z].
16. A. Matlock, J. Zhu, and L. Tian, "Multiple-scattering simulator-trained neural network for intensity diffraction tomography," Opt. Express, OE **31**(3), 4094–4107, Optica Publishing Group (2023) [doi:10.1364/OE.477396].
17. P. Ledwig and F. E. Robles, "Epi-mode tomographic quantitative phase imaging in thick scattering samples," Biomed. Opt. Express, BOE **10**(7), 3605–3621, Optica Publishing Group (2019) [doi:10.1364/BOE.10.003605].
18. P. Ledwig and F. E. Robles, "Quantitative 3D refractive index tomography of opaque samples in epi-mode," Optica, OPTICA **8**(1), 6–14, Optica Publishing Group (2021) [doi:10.1364/OPTICA.410135].
19. F. Eadie et al., "Fourier ptychography microscopy for digital pathology," Journal of Microscopy **300**(2), 260–285 (2025) [doi:10.1111/jmi.70001].
20. T. H. Nguyen et al., "Gradient light interference microscopy for 3D imaging of unlabeled specimens," Nat Commun **8**(1), 210, Nature Publishing Group (2017) [doi:10.1038/s41467-017-00190-7].
21. W. Krauze et al., "3D scattering microphantom sample to assess quantitative accuracy in tomographic phase microscopy techniques," 1, Sci Rep **12**(1), 19586, Nature Publishing Group (2022) [doi:10.1038/s41598-022-24193-7].
22. P. Ossowski et al., "Near-infrared, wavelength, and illumination scanning holographic tomography," Biomed. Opt. Express, BOE **13**(11), 5971–5988, Optica Publishing Group (2022) [doi:10.1364/BOE.468046].
23. E. Hartveit, Ed., *Multiphoton Microscopy*, Springer, New York, NY (2019) [doi:10.1007/978-1-4939-9702-2].
24. Y. Zhang et al., "3D imaging of optically cleared tissue using a simplified CLARITY method and on-chip microscopy," 8, Science Advances **3**(8) (2017) [doi:10.1126/sciadv.1700553].
25. M. Rogalski et al., "Gigavoxel-Scale Multiple-Scattering-Aware Lensless Holotomography," arXiv:2508.00567, arXiv (2025) [doi:10.48550/arXiv.2508.00567].
26. B. Mirecki et al., "Low-intensity illumination for lensless digital holographic microscopy with minimized sample interaction," Biomedical Optics Express **13**(11), 5667–5682, Optica Publishing Group (2022).
27. M. Rogalski et al., "Hybrid Iterating-Averaging Low Photon Budget Gabor Holographic Microscopy," ACS Photonics **12**(4), 1771–1782, American Chemical Society (2025) [doi:10.1021/acsphotonics.4c01863].
28. R. Gerchberg, "A practical algorithm for the determination of plane from image and diffraction pictures," Optik **35**(2), 237–246 (1972).
29. A. Greenbaum, U. Sikora, and A. Ozcan, "Field-portable wide-field microscopy of dense samples using multi-height pixel super-resolution based lensfree imaging," Lab Chip **12**(7), 1242 (2012) [doi:10.1039/c2lc21072j].

30. Y. Rivenson et al., "Sparsity-based multi-height phase recovery in holographic microscopy," Sci Rep **6**(1), 37862 (2016) [doi:10.1038/srep37862].
31. L. Hacker et al., "Criteria for the design of tissue-mimicking phantoms for the standardization of biophotonic instrumentation," Nat. Biomed. Eng **6**(5), 541–558, Nature Publishing Group (2022) [doi:10.1038/s41551-022-00890-6].
32. R. Khir, W. Bachir, and F. S. Ismael, "Pasteurized milk vs intralipid 20% for scattering-based optical phantoms: a comparative study," J Opt **53**(1), 720–727 (2024) [doi:10.1007/s12596-023-01508-z].
33. J. Bertolotti and O. Katz, "Imaging in complex media," 9, Nat. Phys. **18**(9), 1008–1017, Nature Publishing Group (2022) [doi:10.1038/s41567-022-01723-8].
34. L. Zhu et al., "Large field-of-view non-invasive imaging through scattering layers using fluctuating random illumination," Nat Commun **13**(1), 1447, Nature Publishing Group (2022) [doi:10.1038/s41467-022-29166-y].
35. L. Zhou, Y. Xiao, and W. Chen, "Imaging Through Turbid Media With Vague Concentrations Based on Cosine Similarity and Convolutional Neural Network," IEEE Photonics Journal **11**(4), 1–15 (2019) [doi:10.1109/JPHOT.2019.2927746].
36. P. Arcab et al., "Low-dose Chemically Specific Bioimaging via Deep-UV Lensless Holographic Microscopy on a Standard Camera," arXiv:2511.21311, arXiv (2025) [doi:10.48550/arXiv.2511.21311].
37. W.-P. Hu et al., "Helium–Neon Laser Irradiation Stimulates Cell Proliferation through Photostimulatory Effects in Mitochondria," Journal of Investigative Dermatology **127**(8), 2048–2057 (2007) [doi:10.1038/sj.jid.5700826].
38. P. P. Laissue et al., "Assessing phototoxicity in live fluorescence imaging," Nat Methods **14**(7), 657–661, Nature Publishing Group (2017) [doi:10.1038/nmeth.4344].
39. P. Arcab et al., "Experimental optimization of lensless digital holographic microscopy with rotating diffuser-based coherent noise reduction," Opt. Express **30**(24), 42810 (2022) [doi:10.1364/OE.470860].
40. M. J. Lopera et al., "Lensless Mueller holographic microscopy with robust noise reduction for multiplane polarization imaging," Optics & Laser Technology **181**, 111936 (2025) [doi:10.1016/j.optlastec.2024.111936].
41. A. S. Galande et al., "High-resolution lensless holographic microscopy using a physics-aware deep network," JBO **29**(10), 106502, SPIE (2024) [doi:10.1117/1.JBO.29.10.106502].
42. K. Matsushima and T. Shimobaba, "Band-Limited Angular Spectrum Method for Numerical Simulation of Free-Space Propagation in Far and Near Fields," Opt. Express, OE **17**(22), 19662–19673, Optica Publishing Group (2009) [doi:10.1364/OE.17.019662].

**Supplementary Material: Near-infrared lensless holographic microscopy on a visible sensor enables label-free high-throughput imaging in strong scattering**

Emilia Wdowiak,[a,*] Piotr Arcab,[a] Mikołaj Rogalski,[a] Anna Chwastowicz,[b,c] Paweł Matryba,[b,c] Małgorzata Lenarcik,[d,e] Julianna Winnik,[a] Piotr Zdańkowski,[a] Maciej Trusiak[a,**]

[a] Warsaw University of Technology, Institute of Micromechanics and Photonics, 8 Sw. A. Boboli St., 02-525 Warsaw, Poland

[b] Laboratory of Neurobiology, Nencki Institute of Experimental Biology of Polish Academy of Sciences, Warsaw, Poland

[c] Department of Immunology, Medical University of Warsaw, Warsaw, Poland

[d] Department of Pathology, Maria Sklodowska-Curie National Research Institute of Oncology, Warsaw, Poland

[e] Department of Gastroenterology, Hepatology and Clinical Oncology, Centre of Postgraduate Medical Education, Warsaw, Poland

*Emilia Wdowiak, E-mail: emilia.wdowiak.dokt@pw.edu.pl
**Maciej Trusiak, E-mail: maciej.trusiak@pw.edu.pl

**Supplementary Note 1: Testing LDHM in controlled scattering conditions**

Several configurations were investigated, as illustrated in Figs. S1(a2-a5). In configuration Fig. S1(a2), the scattering layer SL is placed below the target, in direct contact with its lower surface and positioned between the illumination source and the object. Configuration Fig. S1(a3) is similar, but the SL is separated from the target by a 10 mm gap. In configuration Fig. S1(a4), the SL is placed above the target, between the sample and the detector, in direct contact with the upper surface of the target. Configuration in Fig. S1(a5) employs SLs both upstream and downstream of the target. In all configurations, the total thickness of the scattering medium was fixed at 200 μm. In configuration Fig. S1(a5), this total thickness was realized using two 100 μm layers placed on either side of the target. To approximate the RI matching and immersion conditions of configuration Fig. S1(a1) in configurations Figs. S1(a2-a5), the test target was immersed in distilled water (RI ≈ 1.331 at 20°C and $\lambda = 632$ nm[1]), while the SL consisted of whole milk (RI ≈ 1.347 under the same conditions[2]). Water was assumed to be non-scattering, serving solely as an RI-matching immersion medium.

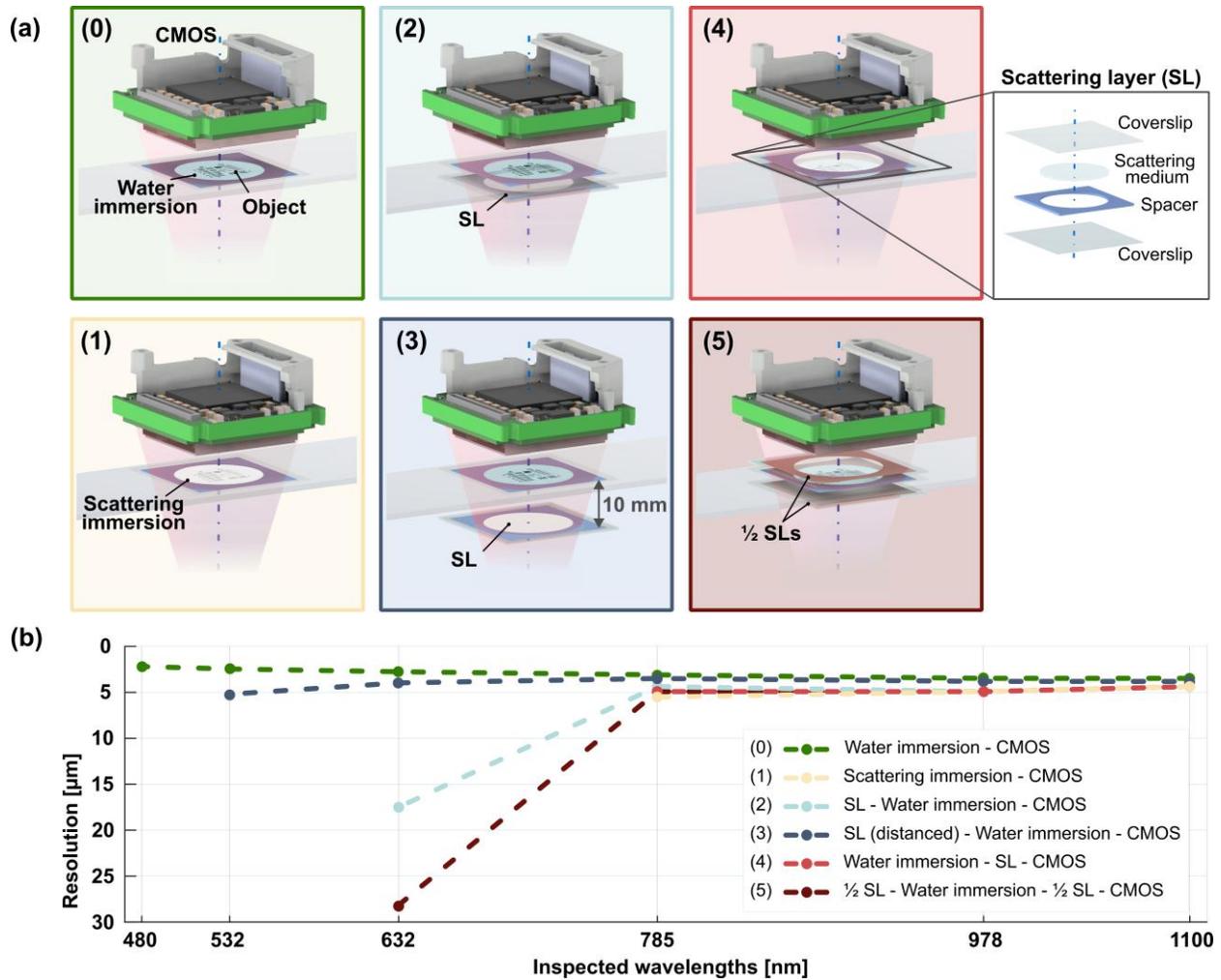

**Fig. S1** LDHM system and investigated scattering configurations. (a0) Water immersion without scattering. (a1) Milk immersion representing realistic biological scattering conditions. (a2) SL positioned below the target and in direct contact with the lower surface of the test. (a3) Same as (a2), with a 10 mm separation between the SL and the target. (a4) SL positioned above the target, between the sample and the detector, in direct contact with the upper surface of the test. (a5) SL positioned both above and below the target. (b) Lateral resolution as a function of scattering configuration for six illumination wavelengths spanning the VIS and NIR. Measured resolution values (colored data points) are connected by straight lines to guide the eye. Missing data points indicate imaging conditions under which the target features were not resolvable.

To determine whether any of the alternative configurations could serve as a practical replacement for the fully immersed configuration Fig. S1(a1) we fabricated a custom phase USAF-resembling resolution target using two-photon polymerization (TPP). This technique enables highly localized polymerization of a photosensitive resin via nonlinear two-photon absorption, allowing three-dimensional (3D) micro- and nanostructures to be produced with submicron precision. Fabrication was carried out using a commercial TPP system (Photonic Professional GT2, Nanoscribe GmbH), equipped with a 63× oil-immersion objective (NA = 1.4) and galvanometric beam scanning. The photoresist IP-Dip 2 (Nanoscribe GmbH) was employed as the printing material, yielding a refractive index of approximately 1.547 after full polymerization.

The target geometry was derived from a binary USAF resolution pattern. During design preparation, the binary intensity values of the USAF image were mapped to laser exposure levels, and the two-dimensional pattern was extended along the axial (z) direction to form a three-dimensional structure. File preparation and fabrication control were performed using DeScribe software (Nanoscribe GmbH). Figure S2(a) shows the input USAF pattern, while Fig. S2(b) presents a 3D visualization of the corresponding print file.

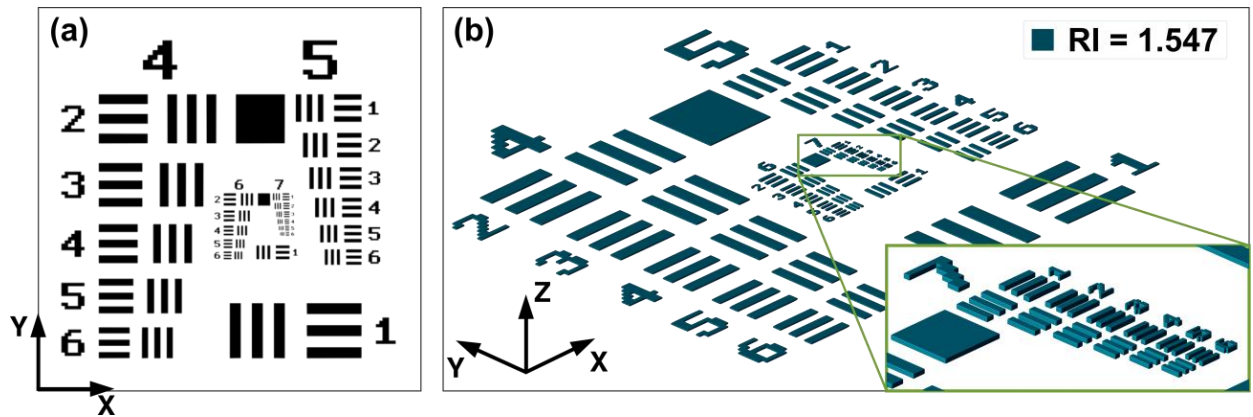

**Fig. S2** Fabrication and characterization of the custom phase USAF-resembling target fabricated with TPP. (a) Binary USAF resolution pattern used as the input design. (b) 3D visualization of the geometry print file.

After fabrication, the printed structure was developed by sequential immersion in isopropyl alcohol for 10 min to remove unpolymerized resin, followed by rinsing in propylene glycol methyl ether acetate (PGMEA), and then air drying. The structure was printed on a fused silica substrate (25 mm × 25 mm × 0.7 mm). The designed feature height of the phase target was 2 μm, corresponding to an expected phase delay (under water or milk immersion) of approximately 4.6 rad at 480 nm and 2 rads at 1100 nm illumination. Representative phase reconstructions of the fabricated target (groups 6-7) are shown in Fig. S3.

While this target might not match the fabrication precision of high-quality commercial resolution standards, it can be immersed in milk and treated as a disposable sample, making it suitable for comparative experiments. The target was first measured under water immersion alone (configurations in Figs. S1(a0, a2-a5)), then dried, and followed by milk immersion (configuration in Fig. S1(a1)). All scattering configurations described in Figs. S1(a0-a5) were therefore evaluated using the same test target to ensure consistency. For each configuration, the smallest resolvable features were determined using the LDHM system across all six investigated illumination wavelengths.

Resolution performance obtained from configurations Figs. S1(a2-a5) was then directly compared against the reference immersed configuration Fig. S1(a1). The configuration which resolution readings most closely matched that of configuration Fig. S1(a1) was considered a viable substitute. The resulting resolution as a function of scattering configuration is summarized in Fig. S1(b). These results show that configuration Fig. S1(a4) - in which the SL is placed between the sample and the detector, exhibits resolution performance nearly identical to that of the fully immersed configuration Fig. S1(a1).  Reconstructed results for representative wavelengths (532 nm and 978 nm) are provided in Fig. S3. Based on this finding, all subsequent experiments in this study were performed using configuration Fig. S1(a4), allowing the use of high-quality commercial resolution targets without direct exposure to any scattering/immersion medium.

Notably, configuration Fig. S1(a3), in which the SL is positioned below the target but separated by a 10 mm gap, exhibited improved resolution compared to configuration Fig. S1(a2), where the SL is in direct contact with the target. This observation suggests that increasing the distance between the SL and the object can partially mitigate resolution degradation under multiple scattering conditions. Moreover, in the NIR range, configuration Fig. S1(a3) yields results approaching those of configuration Fig. S1(a0), corresponding to the no-scattering case with water immersion only. On the other hand, configurations with the SL placed directly before the object - Fig. S1(a2) - and with the SL distributed symmetrically before and after the sample - Fig. S1(a5) - produce NIR performance close to the ideal scenario represented by Fig. S1(a1). Nevertheless, these configurations remain resolvable even at 632 nm, at the long-wavelength end of the VIS range. Overall, the performance differences across all investigated configurations remain relatively modest with respect to the ideal case, suggesting a high tolerance of the system to scattering geometry. More pronounced distinctions could likely emerge in a more detailed parametric study, for example through systematic variation of the SL thickness.

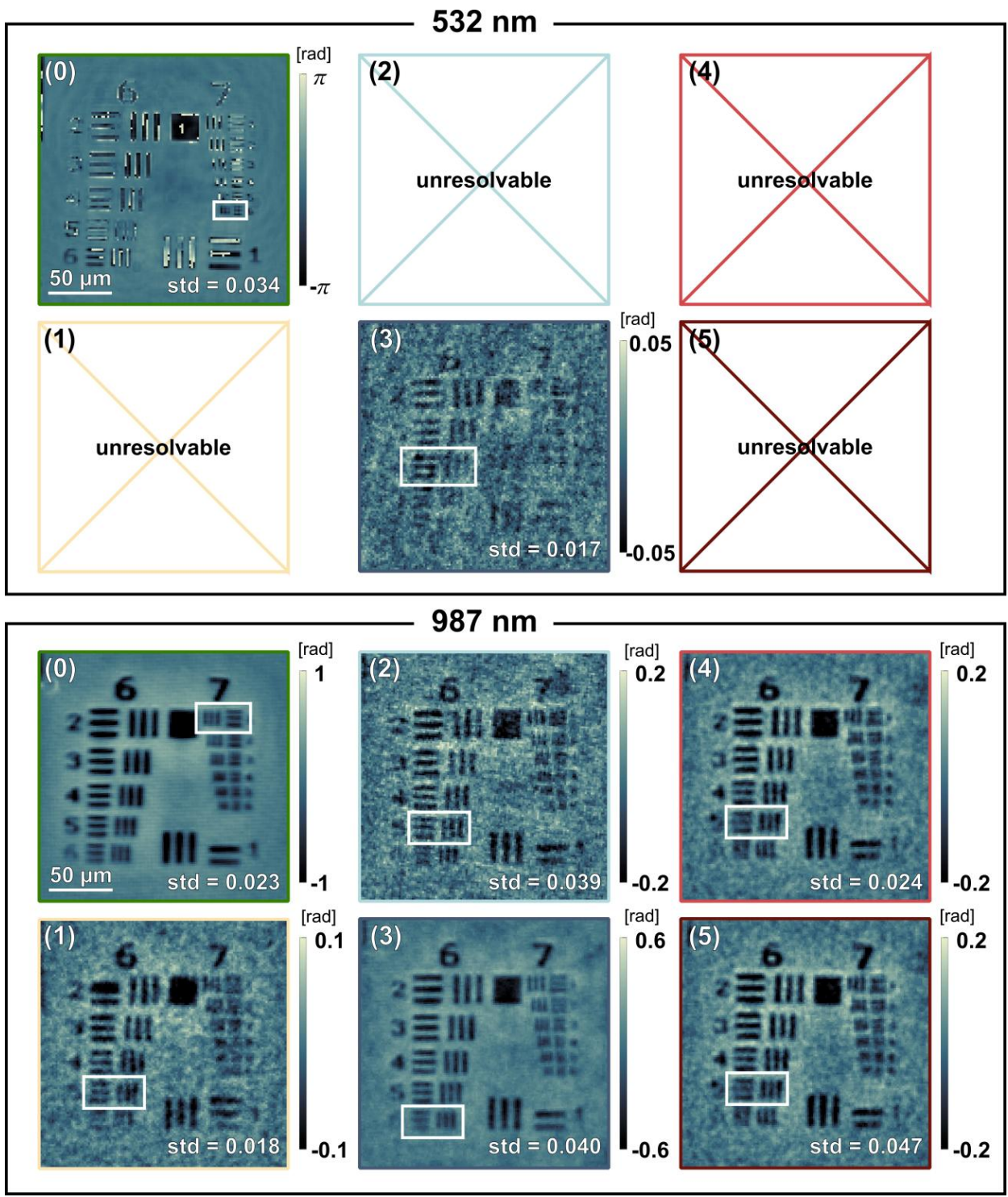

**Fig. S3** Phase reconstruction results for VIS (532 nm) and NIR (978 nm) illumination. The numbering corresponds to the scattering system configurations shown in Fig. S1. White boxes mark the smallest resolvable element identified in each reconstruction.

**Supplementary Note 2: Wavelength-dependent LDHM phase and amplitude resolution results under increasing scattering**

Representative reconstructed amplitude and phase results obtained for selected SL thicknesses are presented to complement the resolution analysis shown in the main manuscript Fig. 3. The reconstructions are provided for SL thicknesses of 130 µm, 230 µm, 520 µm, and 910 µm, illustrated in Figs. S4–S7, respectively, to enable direct visual assessment of image quality across progressively increasing scattering conditions.

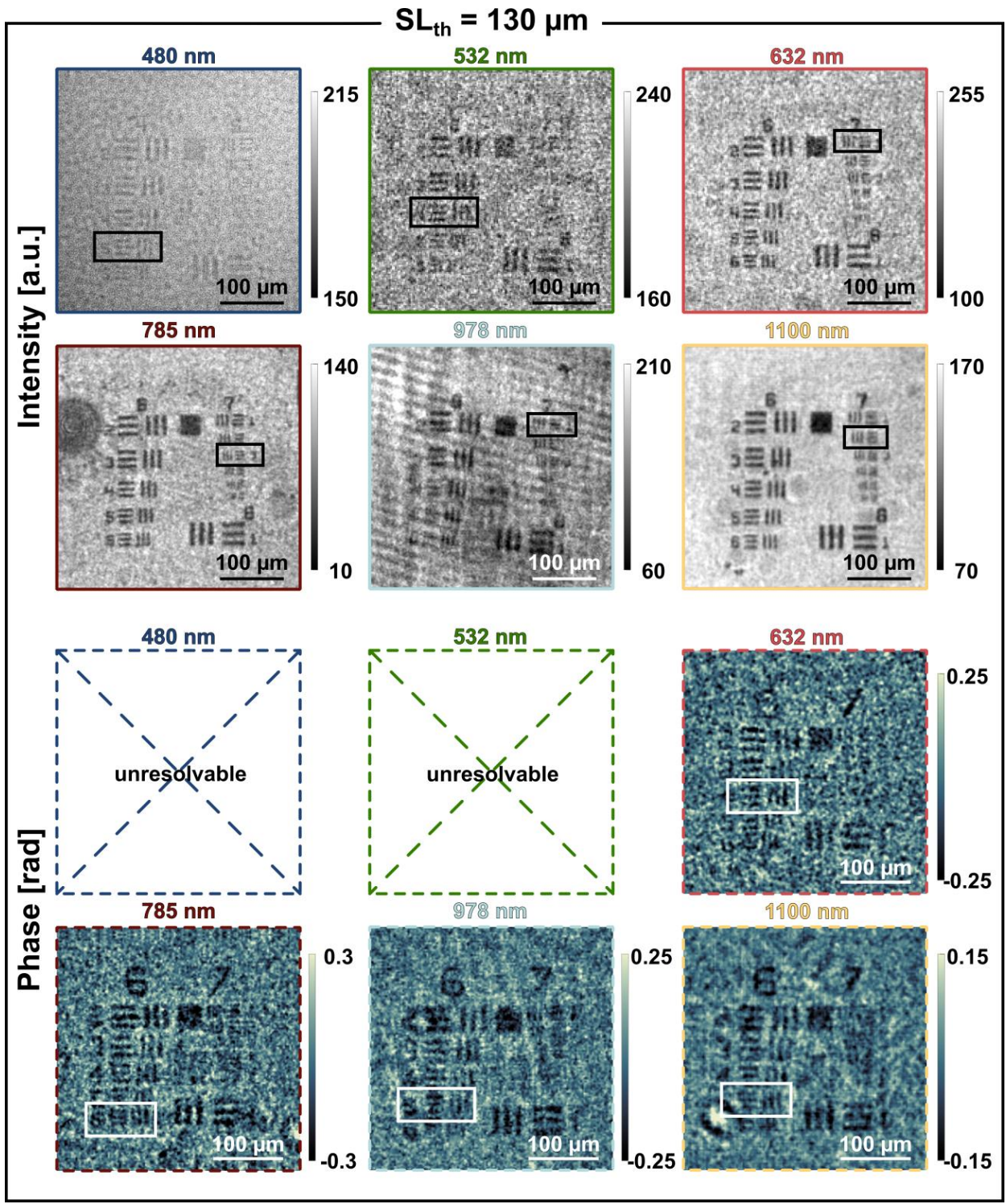

**Fig. S4** Reconstructed amplitude and phase images obtained for an SL thickness of 130 μm under the investigated illumination wavelengths. Black and white boxes indicate the smallest resolvable element identified in each reconstruction.

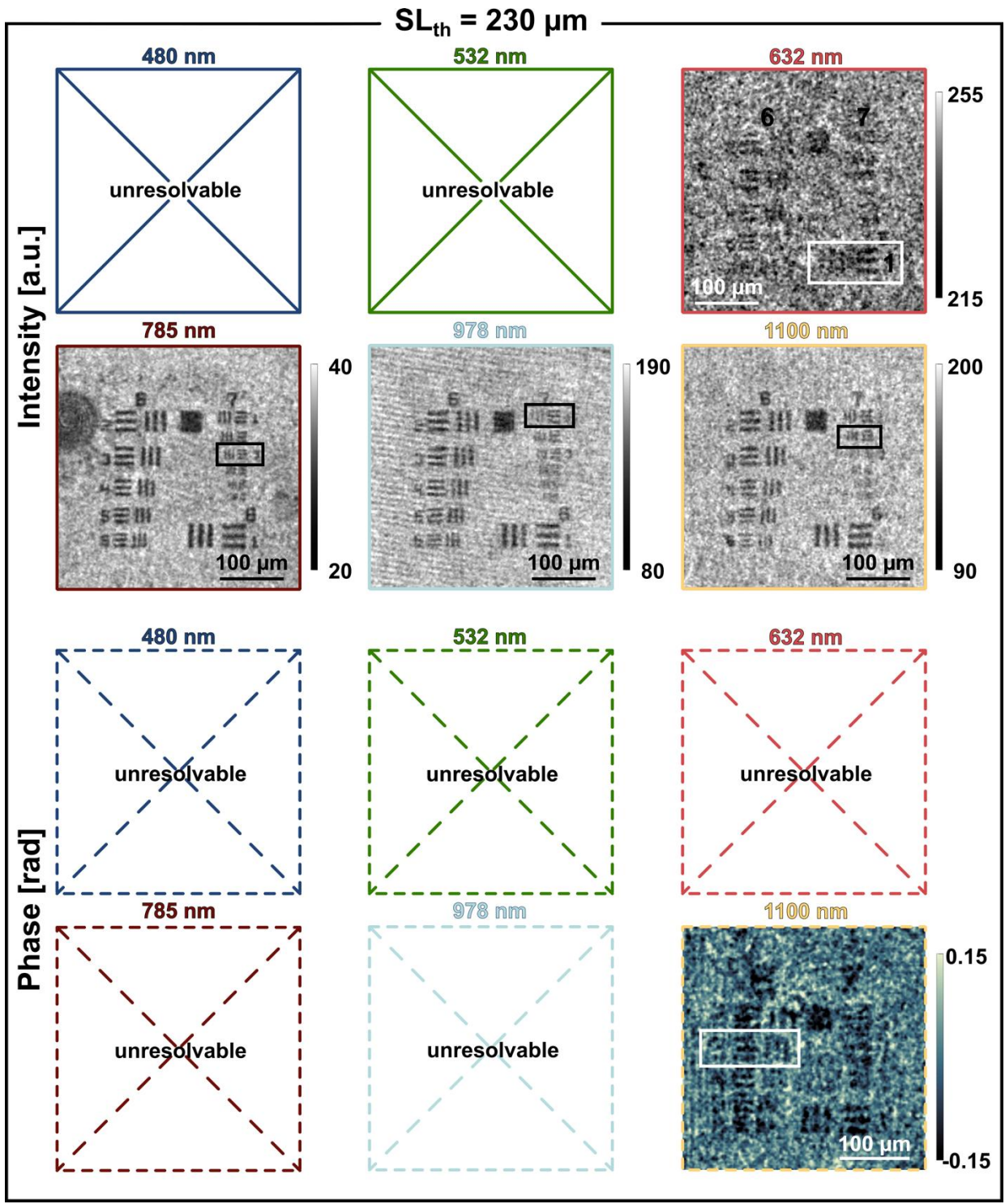

**Fig. S5** Reconstructed amplitude and phase images obtained for an SL thickness of 230 μm under the investigated illumination wavelengths. Black and white boxes indicate the smallest resolvable element identified in each reconstruction.

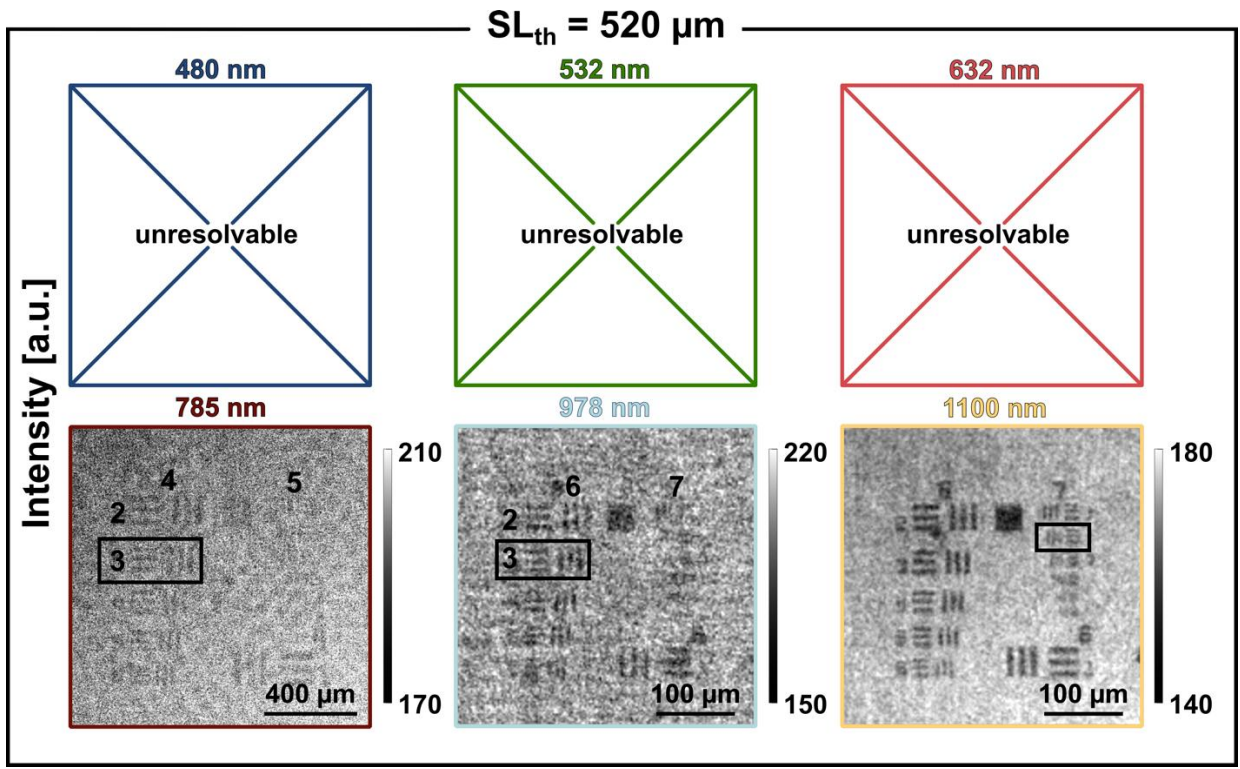

**Fig. S6** Reconstructed amplitude images obtained for an SL thickness of 520 µm under the investigated illumination wavelengths. Black boxes indicate the smallest resolvable element identified in each reconstruction.

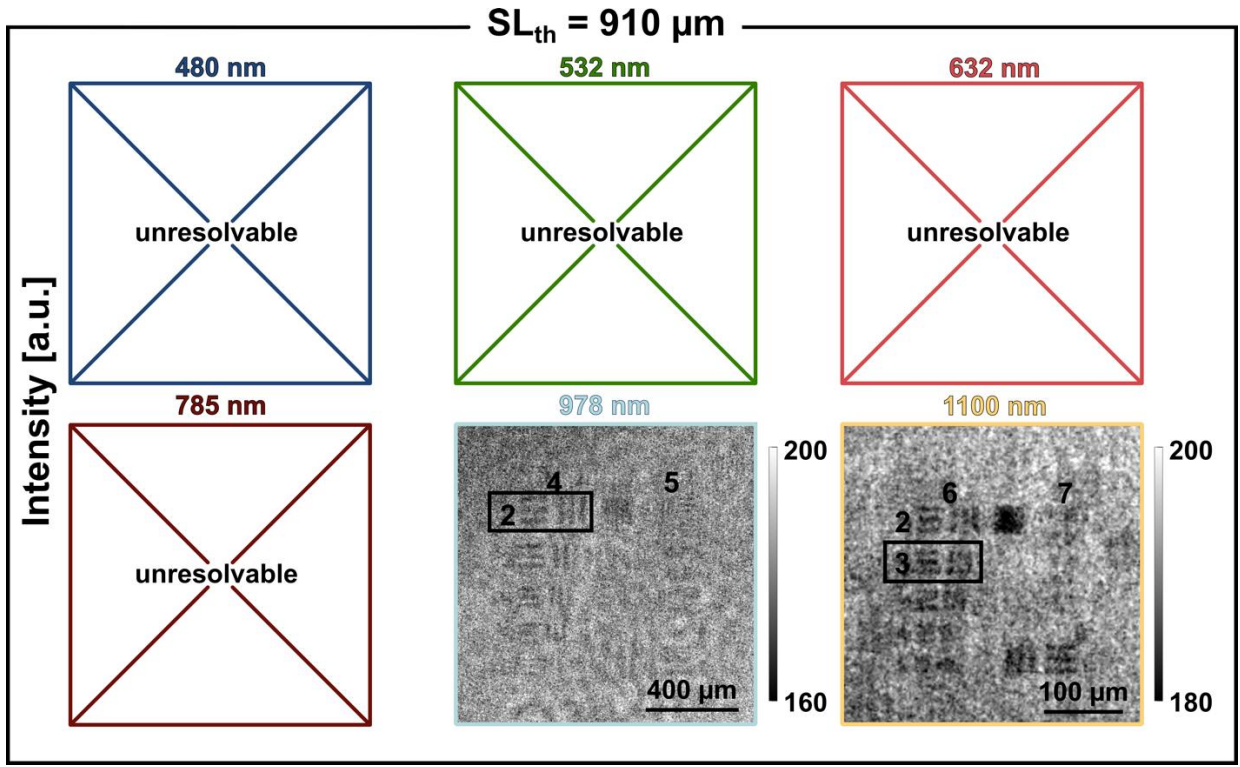

**Fig. S7** Reconstructed amplitude images obtained for an SL thickness of 910 μm under the investigated illumination wavelengths. Black boxes indicate the smallest resolvable element identified in each reconstruction.

**Supplementary Note 3: Visualization of sample-sensor distance effects**

Figure S8 provides representative raw holograms, normalized holograms, and normalized reconstructed amplitude maps corresponding to the amplitude resolution study discussed in the main manuscript. Images were normalized by dividing each frame by its heavily Gaussian-blurred version ($\sigma$ = 60 px). The data span sample-sensor propagation distances from 2.85 mm to 12.80 mm under strong scattering conditions (300 μm-thick SL). As shown in the raw holograms, a substantial reduction in detected intensity is observed with increasing propagation distance, consistent with geometric spreading and scattering-induced attenuation. After normalization, the holograms reveal progressively enhanced visibility and contrast of the resolution target features as the propagation distance increases. This improvement is further reflected in the normalized reconstructed amplitude maps, where an increase in SNR and clearer definition of fine features is observed at larger distances. These qualitative observations visually corroborate the quantitative trends reported in Fig. 4 from the main manuscript, confirming that increased sample-sensor distance can improve reconstruction quality in strongly scattering regimes despite reduced overall signal intensity.

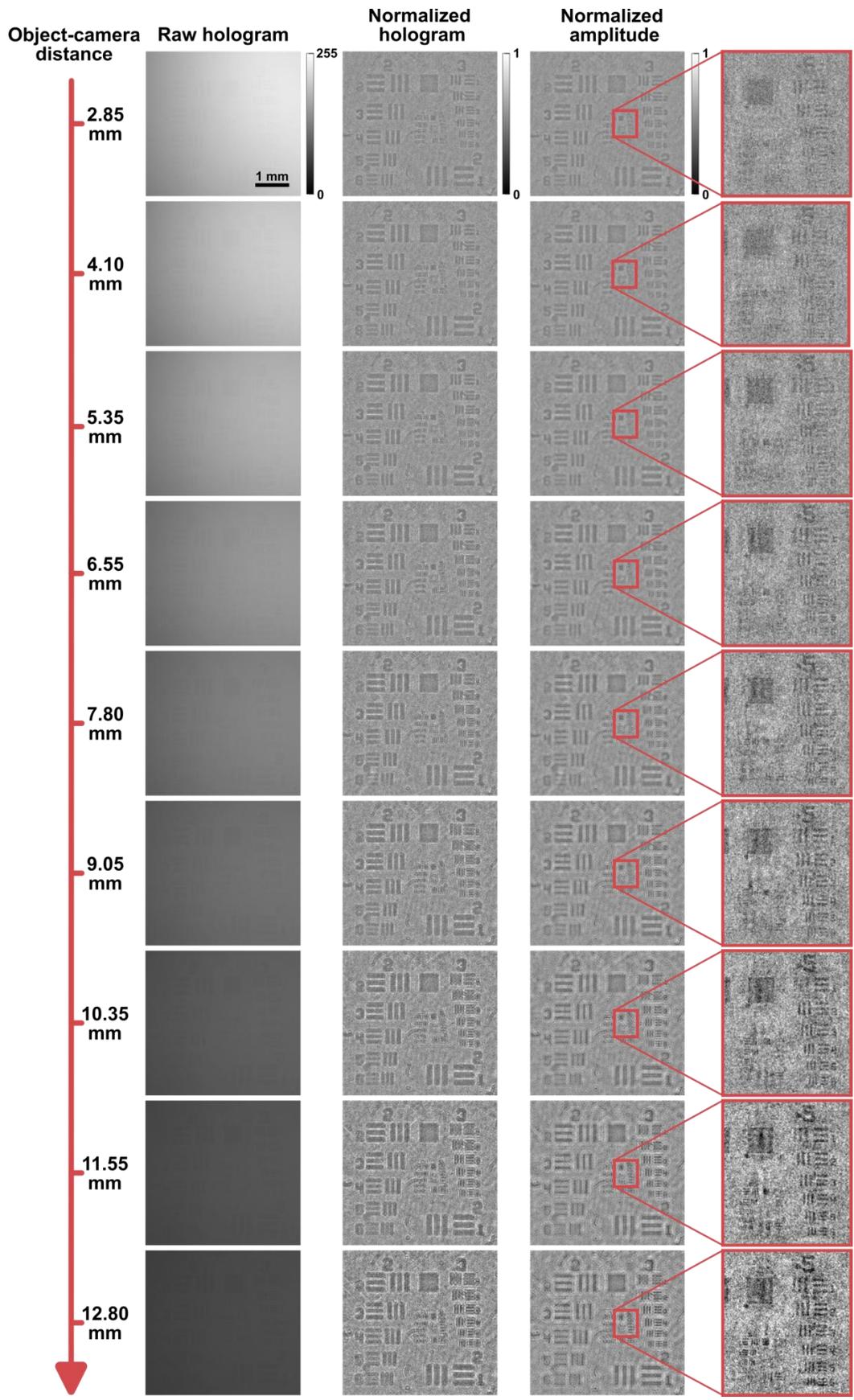
37

**Fig. S8** Raw holograms of the amplitude USAF target acquired at increasing sample-sensor distances (from 2.85 mm to 12.8 mm) with 632 nm illumination, 300 μm-thick SL, corresponding normalized holograms, and normalized single-frame-reconstructed amplitude maps with enlarged regions of interest.

Figure S9 shows the corresponding multi-frame GS normalized amplitude reconstruction obtained from the same dataset presented in Fig. S8, using nine holograms acquired over the initial portion of the studied sample-sensor distance range (2.85 - 12.80 mm). This distance interval corresponds to the regime in which the most pronounced resolution improvement is observed, as shown in Fig. 4 of the main manuscript; beyond approximately 14 mm, the resolution begins to degrade. The reconstruction was performed under strong scattering conditions (300 μm SL, 632 nm illumination). Compared with the individual single-frame reconstructions in Fig. S8, the GS reconstruction exhibits substantially reduced noise and a smoother background, confirming the noise-suppression benefit of multi-frame processing. At the same time, the effective resolution of the GS result reflects an average performance across the selected propagation distances, rather than matching the best resolution achieved by the most favorable single frame.

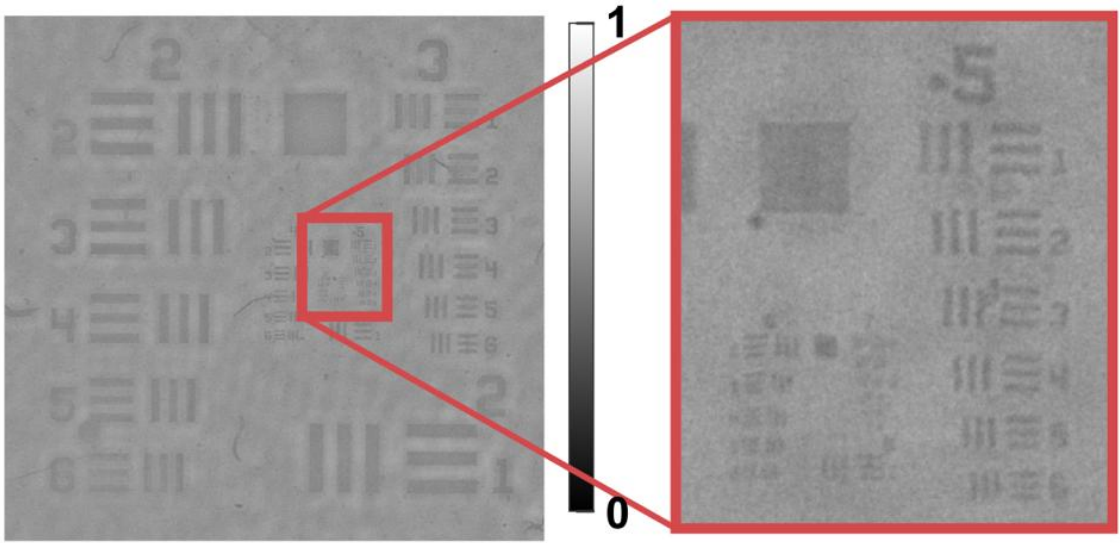

**Fig. S9** GS normalized amplitude reconstruction of the amplitude USAF test under strong scattering conditions (300 μm-thick SL), illuminated with 632 nm wavelength. The multi-frame reconstruction is performed with nine single holograms presented in the Fig. S2.

This trend is consistent with observations from the controlled scattering configuration study (Fig. S1), where configuration Fig. S1(a3), in which the SL is separated from the target by a 10 mm gap, yielded improved resolution compared with the configuration in which the SL was in direct contact with the target. This behavior is consistent with the van Cittert-Zernike theorem, which predicts that spatially incoherent source develops higher coherence upon propagation[3]. In this context, the SL can be viewed as a secondary, spatially incoherent source, and the additional propagation distance allows the field to become more spatially coherent at the object plane. Together, these results indicate that increasing the distance - either between the object and the detector or between the SL and the object - can partially mitigate scattering-induced degradation.

**Supplementary Note 4:** *In vitro* **biological specimens' preparation**

Murine brains and livers were obtained from inbred C57BL/6J mice (bred at the Center for Experimental Medicine in Białystok and the M. Mossakowski Medical Research Center in Warsaw). The animals were provided a controlled environment (temperature 24°C, 12/12 light/dark cycle) with ad libitum access to water and feed. The animals were euthanized by cervical dislocation following exposure to isoflurane (FDG9623, Baxter). Isolated tissues were fixed for 24 h in a 4% paraformaldehyde solution (BD Cytofix/Cytoperm, No. 554722, BD Biosciences) at 4°C before being used in experiments. All procedures were conducted in accordance with the Directive of the European Parliament and Council No. 2010/63/EU on the protection of animals used for scientific purposes. The fixed organs were embedded in a 3% (w/v) agarose solution (Sigma–Aldrich) in water and subsequently sectioned to the desired thickness using a Vibratome (VT1000S, Leica). The sections were stored in 1× PBS (Sigma–Aldrich) supplemented with 0.05% sodium azide (NaN3, Sigma–Aldrich). For imaging, tissue slices were transferred onto a 1 mm-thick microscope slide, immersed in PBS, and covered with a 170 µm-thick coverslip for LDHM measurements.

Figures S10–S14 present enlarged views of brain tissue VIS- and NIR-LDHM results derived from Fig. 6 of the main manuscript to facilitate detailed visualization.

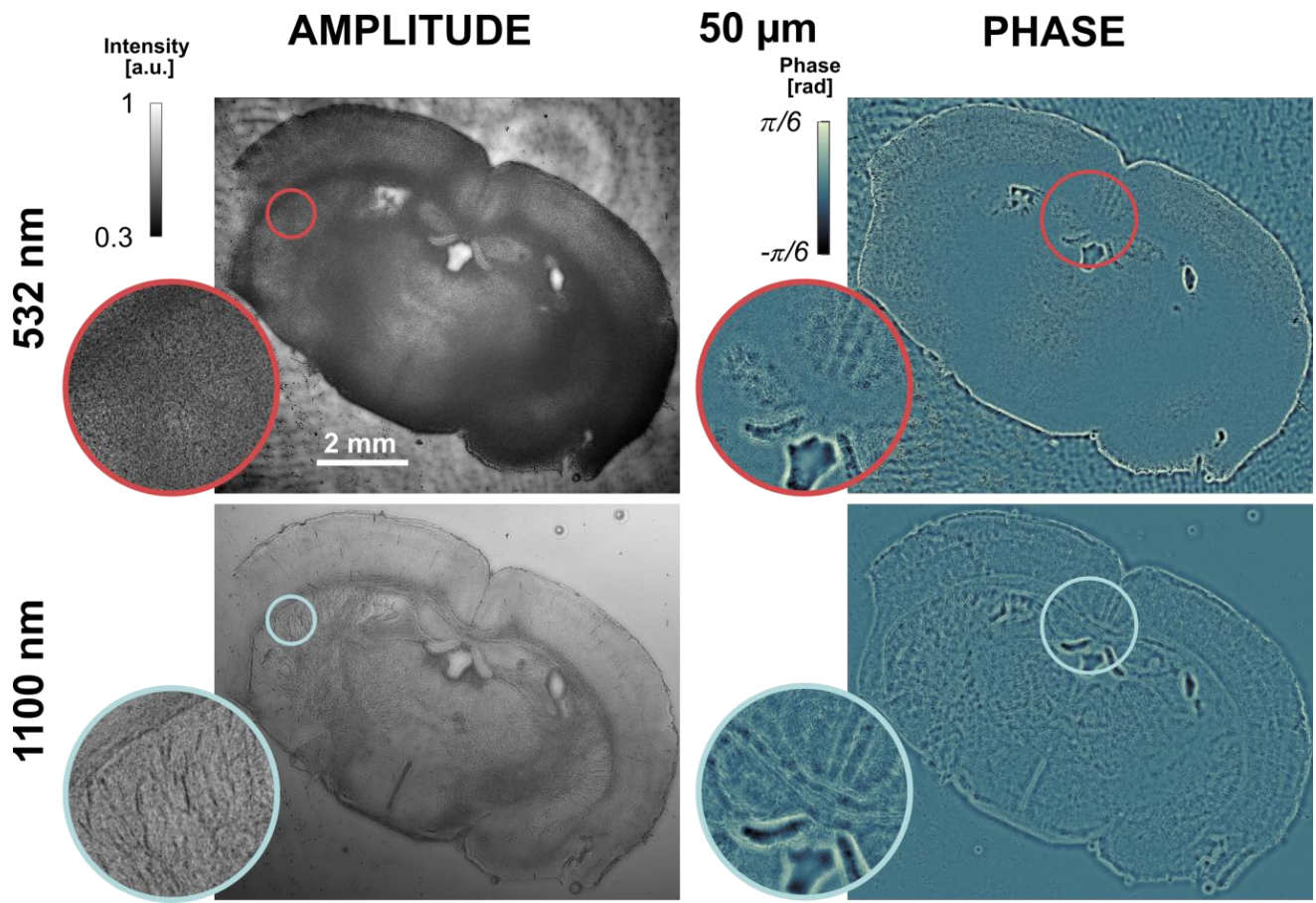

**Fig. S10** Brain tissue imaging using VIS- and NIR-LDHM - 50 μm-thick slice enlarged, taken from Fig. 6 in the main manuscript.

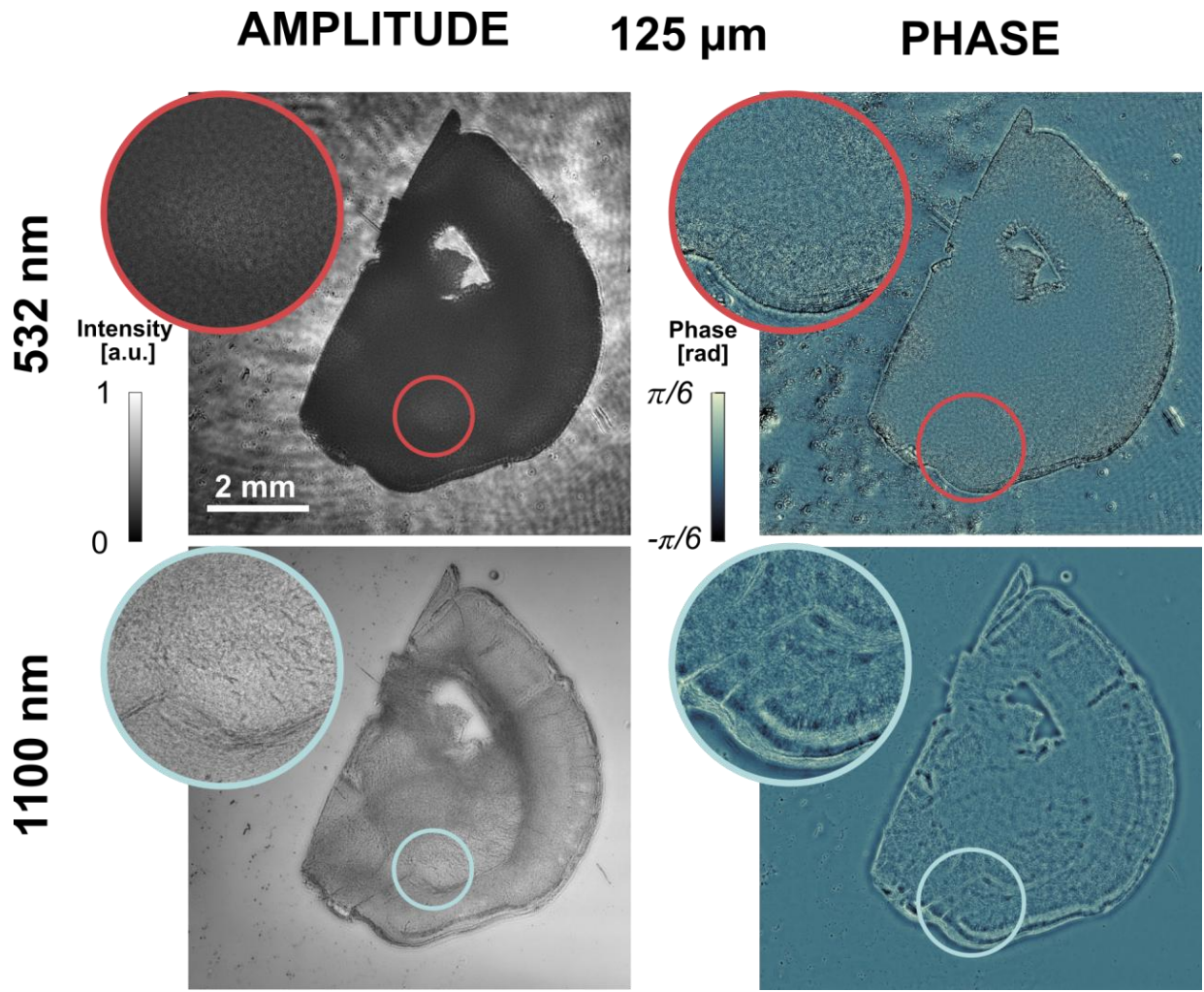

**Fig. S11** Brain tissue imaging using VIS- and NIR-LDHM - 125 μm-thick slice enlarged, taken from Fig. 6 in the main manuscript.

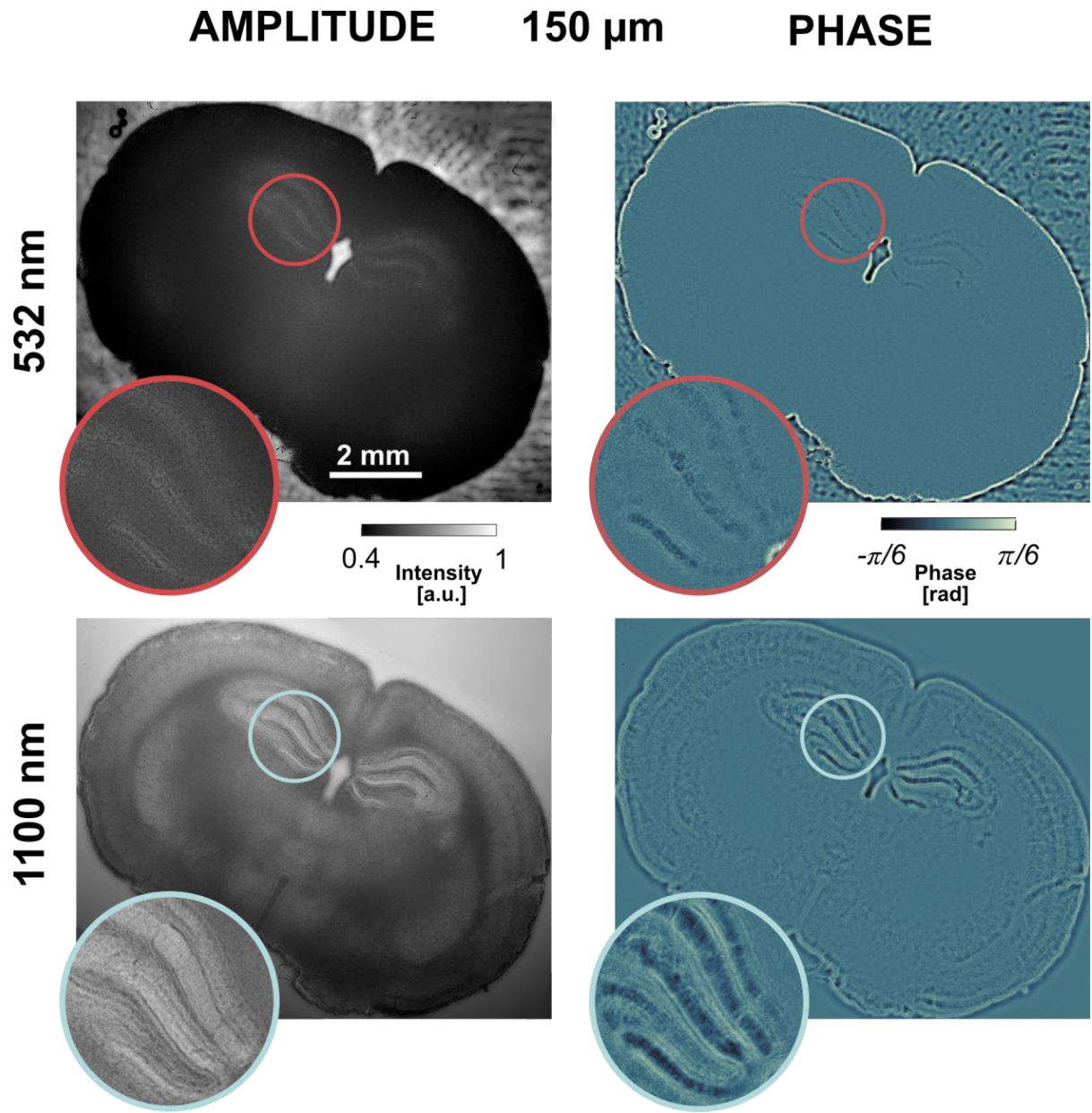

**Fig. S12** Brain tissue imaging using VIS- and NIR-LDHM - 150 μm-thick slice enlarged, taken from Fig. 6 in the main manuscript.

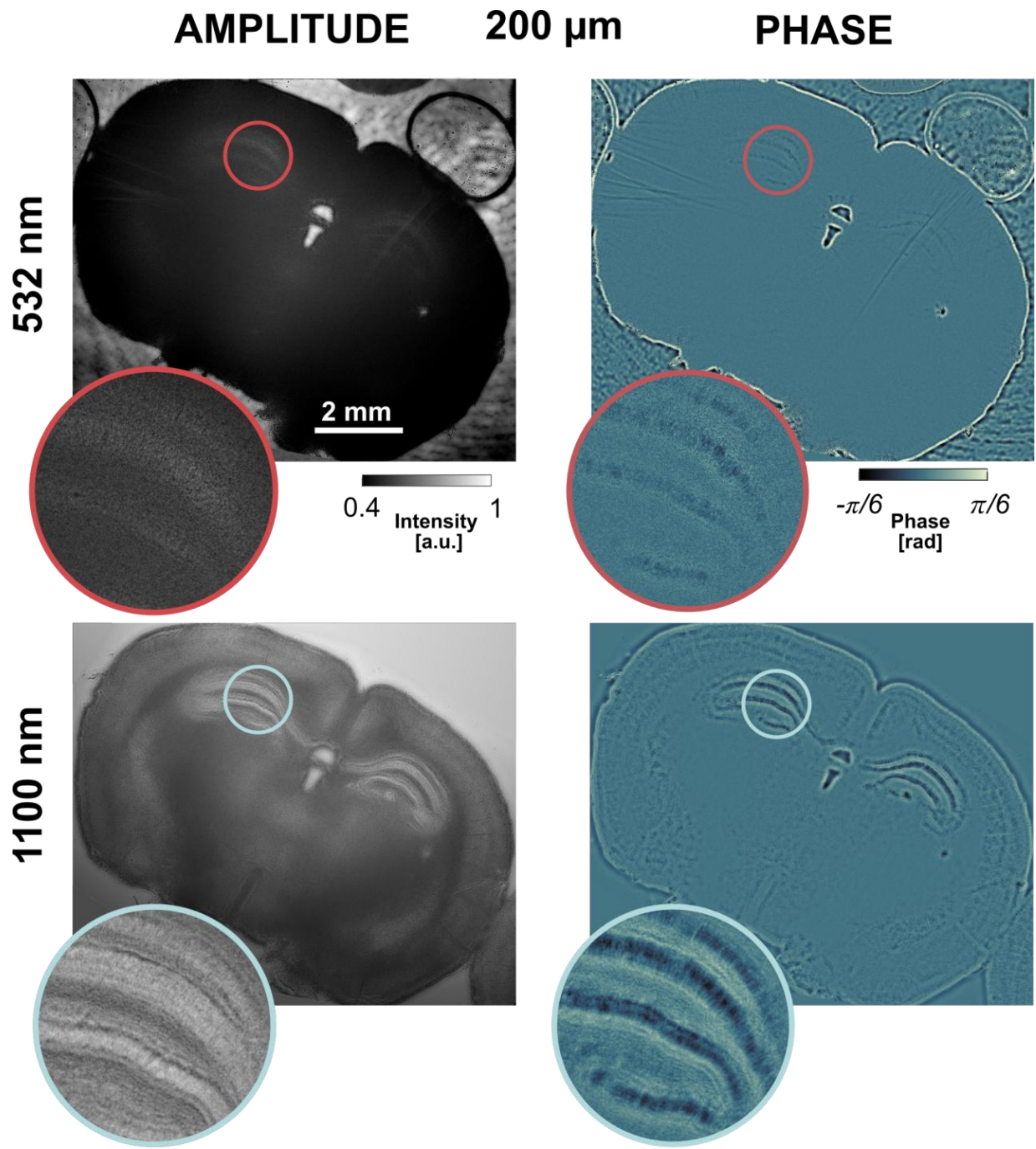

**Fig. S13** Brain tissue imaging using VIS- and NIR-LDHM - 200 μm-thick slice enlarged, taken from Fig. 6 in the main manuscript.

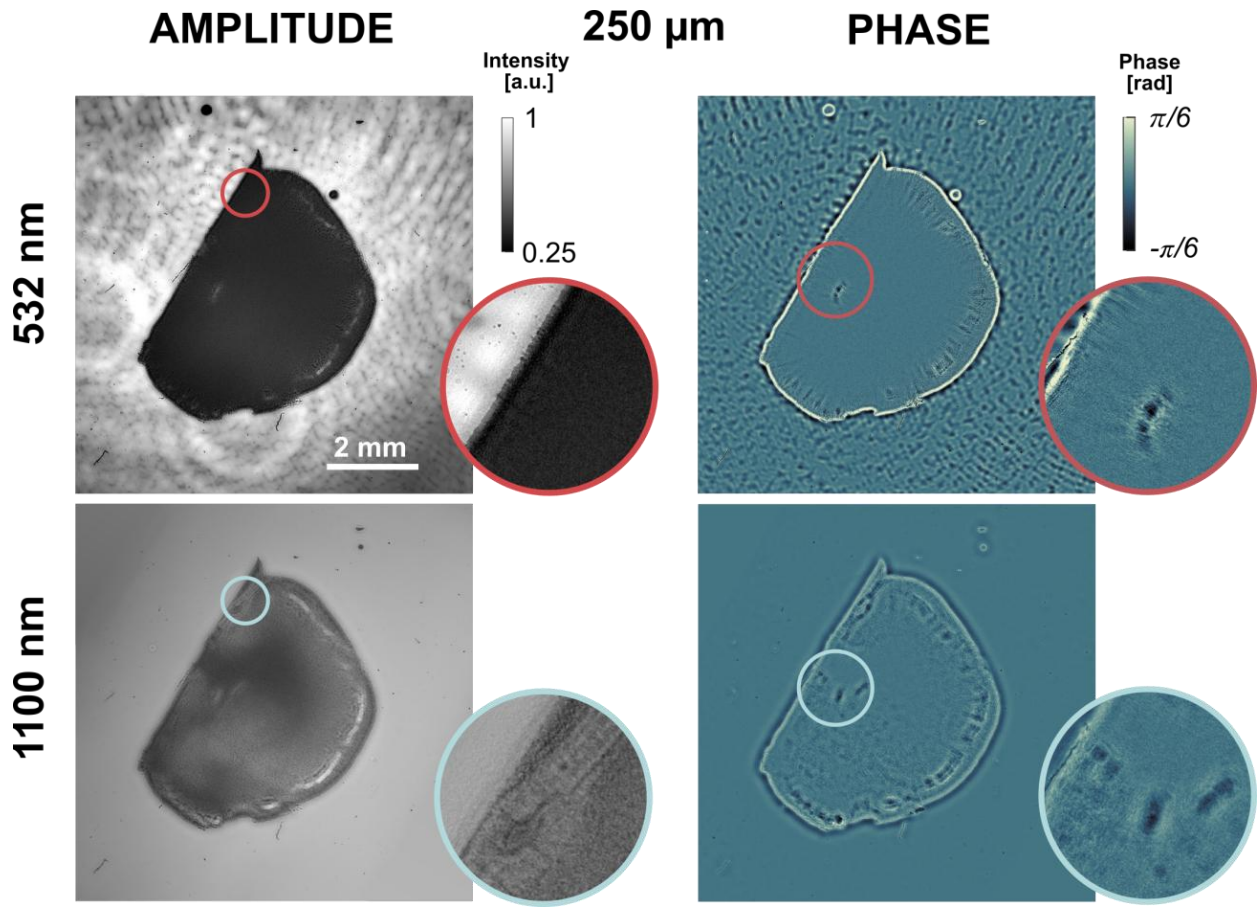

**Fig. S14** Brain tissue imaging using VIS- and NIR-LDHM - 250 μm-thick slice enlarged, taken from Fig. 6 in the main manuscript.

## Supplementary Note 5: Visualization of wavelength-dependent phase imaging under varying scattering level

Figure S15 presents representative GS phase reconstructions of the quantitative phase USAF target for the six illumination wavelengths investigated in this study, acquired under three scattering conditions corresponding to 0 μm, 50 μm, and 100 μm SL thicknesses. These data directly supplement the resolution trends reported in Fig. 3 of the main manuscript by providing visual examples of phase reconstruction quality across wavelength and selected scattering strength levels. As expected, phase imaging performance depends strongly on both parameters. For shorter wavelengths (e.g., 480 nm), phase contrast rapidly degrades with increasing SL thickness, and phase features become unresolvable at 100 μm SL. In contrast, at longer wavelengths - particularly 1100 nm - phase features remain recognizable with only a modest reduction in resolution at the same scattering thickness. Figure S15 also highlights the wavelength-dependent phase modulation imposed by the fixed-height target, with the phase contrast at 1100 nm being more than two times smaller than that at 480 nm, as discussed in the main manuscript. Additionally, beginning at 785 nm, weak parasitic interference fringes become visible, and these artifacts are especially pronounced at 978 nm; their origin and impact are discussed in a subsequent section of the Supplementary Material. Overall, Fig. S15 visually confirms that, despite reduced phase modulation and detector quantum efficiency, NIR illumination enables practically useful phase imaging under scattering conditions, where VIS illumination fails.

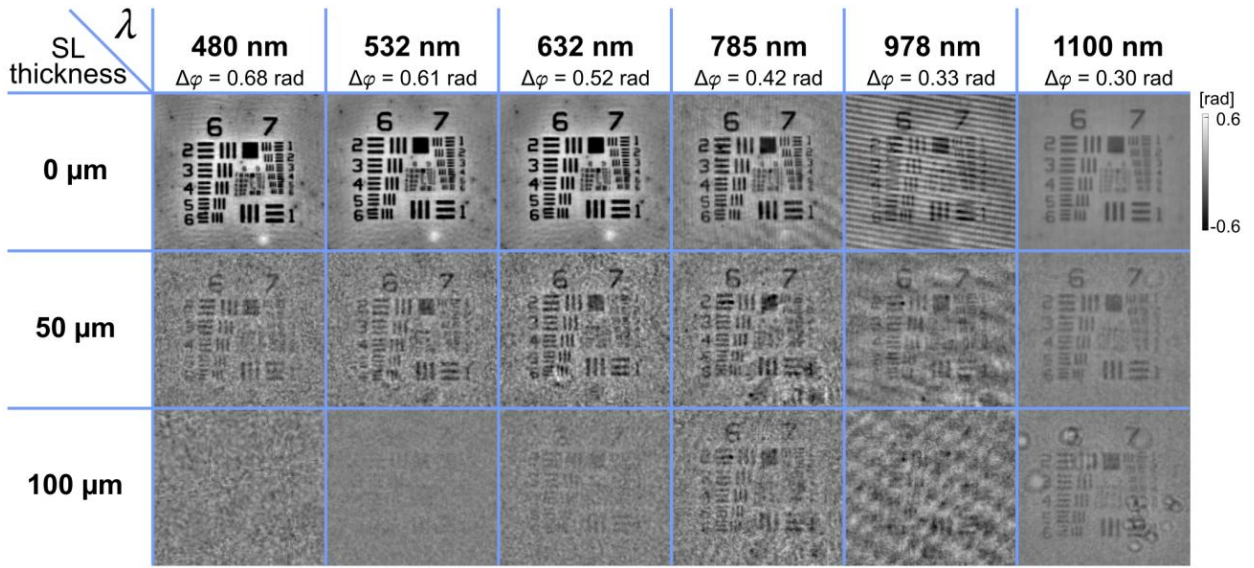

**Fig. S15** GS phase reconstructions of the quantitative phase USAF target acquired at six considered illumination wavelengths under three scattering conditions - 0 μm, 50 μm, and 100 μm SL thickness. The indicated phase shift values $\Delta\varphi$ correspond to the wavelength-dependent phase change introduced by the same USAF test when illuminated at different wavelengths.

**Supplementary Note 6: Wavelength-dependent parasitic interference artifacts in NIR holograms**

During the course of these measurements, wavelength-specific parasitic interference effects were observed for illumination at 978 nm, and similarly at 910 nm (wavelength used in preliminary experiments). Representative examples of these parasitic fringe patterns in the recorded holograms (for wavelengths: 795, 910, 978, and 1100 nm) are shown in Fig. S16. In general, low spatial frequency fringes are commonly observed across all investigated wavelengths and arise from weak interference between reflections at the upper and lower surfaces of the test target. In contrast, for 910 and 978 nm illumination wavelengths, distinctive additional fringe patterns were detected even in the absence of any sample in the optical path. We attribute these parasitic fringes to interference originating within the protective glass layer of the CMOS sensor, where wavelength-dependent internal reflections can produce weak etalon-like effects.

Notably, illumination at 785, 910, and 978 nm was provided by laser diodes - sources with relatively high coherence, which can promote stronger and more pronounced interference fringes than those observed for VIS wavelengths and for 1100 nm illumination generated by the supercontinuum source. The recorded data further indicate the presence of multiple fringe groups: in the absence of an object, fringes are mainly associated with the protective glass layer of the sensor, whereas in the presence of the object and SL, additional interference contributions arise from reflections at the object and within the SL.

These artifacts manifest in the resolution data as increased variability in the 978 nm curves shown in Fig. 3 from the main manuscript (light blue), affecting both amplitude and phase measurements. Specifically, local non-monotonic fluctuations appear as a function of scattering-layer thickness, consistent with constructive or destructive interference contributions that vary with propagation conditions. These parasitic effects do not alter the overall wavelength-dependent trends discussed above and are therefore treated as a secondary experimental artifact. Because the fringe strength at 978 nm varies with SL thickness - being enhanced under some conditions and suppressed under others - we did not attempt numerical removal, in order to process all datasets consistently across wavelengths.

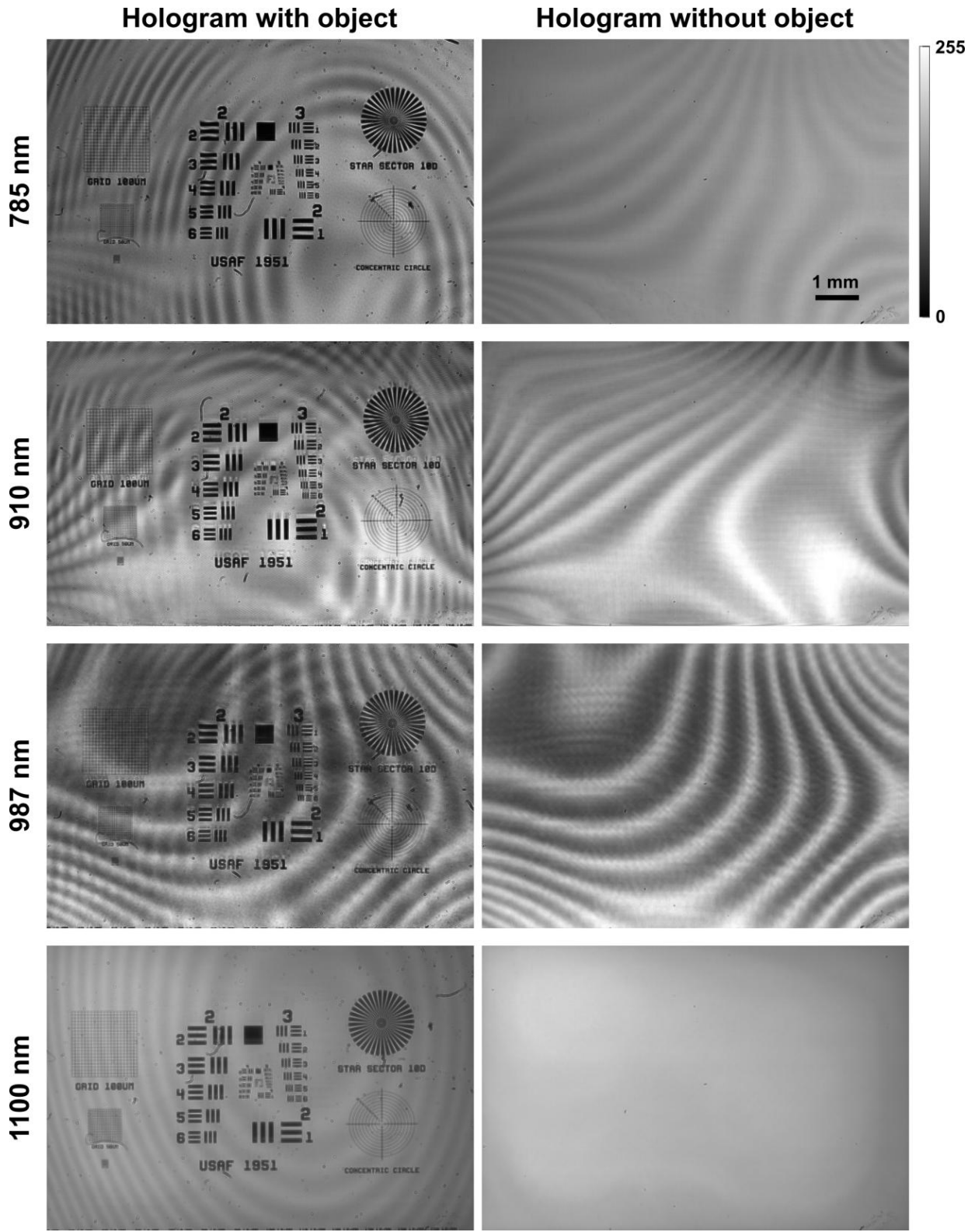

**Fig. S16** Raw holograms with and without amplitude test target acquired at 795 nm, 910 nm, 978 nm, and 1100 nm, showing wavelength-dependent parasitic interference fringes.

*References*

1. M. Daimon and A. Masumura, "Measurement of the refractive index of distilled water from the near-infrared region to the ultraviolet region," Appl. Opt. **46**(18), 3811 (2007) [doi:10.1364/AO.46.003811].
2. A. J. Jääskeläinen, K. Peiponen, and J. A. Räty, "On Reflectometric Measurement of a Refractive Index of Milk," Journal of dairy science **84**, 38–43 (2001) [doi:10.3168/jds.S0022-0302(01)74449-9].
3. "Statistical Optics," in Fundamentals of Photonics, pp. 342–383, John Wiley & Sons, Ltd (1991) [doi:10.1002/0471213748.ch10].